\RequirePackage{fix-cm}
\documentclass{svjour3}       
\smartqed  
\usepackage{graphicx}
\usepackage[margin=1.5in]{geometry}
\usepackage[export]{adjustbox}
\begin{document}

\title{Strong QCD Insights from Excited Nucleon Structure Studies with CLAS and CLAS12}


\author{D.S. Carman \and K. Joo \and V.I. Mokeev}


\institute{D.S. Carman \at 
           Jefferson Laboratory, 12000 Jefferson Ave., Newport News, VA 23602, USA
           \email{carman@jlab.org}
           \and
           K. Joo \at
           University of Connecticut, Storrs, CT 06269, USA
           \email{kyungseon.joo@uconn.edu}
           \and
           V.I. Mokeev \at
           Jefferson Laboratory, 12000 Jefferson Ave., Newport News, VA 23602, USA
           \email{mokeev@jlab.org}}

\date{Received: date / Accepted: date}

\maketitle

\begin{abstract}
Studies of the spectrum of hadrons and their structure in experiments with electromagnetic probes offer unique insight into many facets of the 
strong interaction in the regime of large quark-gluon running coupling, {\it i.e.} the regime of strong QCD. The experimental program within Hall~B 
at Jefferson Laboratory based on data acquired with the CLAS spectrometer using electron and photon beams with energies up to 6~GeV has already 
considerably extended the scope of research in hadron physics in joint efforts between experiment and phenomenological data analysis. Impressive progress 
in relating the hadron structure observables inferred from the data to the strong QCD mechanisms underlying hadron mass generation has been achieved in 
the past decade. These results will be considerably extended with data from the experimental program with the new CLAS12 spectrometer that has begun data 
taking using electron beams with energies up to 11~GeV. With this extended kinematic reach the structure of nucleon resonances will be probed at the 
highest photon virtualities ever achieved in the studies of exclusive electroproduction, which will allow for the exploration of the distance scale where 
$>$98\% of light hadron mass emerges from QCD in the transition of the strong interaction from the regime of quark-gluon confinement to perturbative QCD.
\end{abstract}

\keywords{Excited Nucleon Structure \and Nucleon Resonance Electrocouplings \and Strong QCD \and Emergence of Hadron Mass}

\section{Introduction}
\label{intro}

Significant progress has been realized in studies of excited nucleon state ($N^*$) structure from the data on exclusive meson electroproduction 
measured with the CLAS detector in Hall~B at Jefferson Laboratory (JLab)~\cite{Burkert:2019bhp,Aznauryan:2011qj,Mokeev:2018zxt,Brodsky:2020vco}. The nucleon 
resonance electroexcitation amplitudes ({\it i.e.} the $\gamma_vpN^*$ electrocouplings) have become available for most excited nucleon states 
in the mass range up to 1.8~GeV for photon virtualities $Q^2 < 5$~GeV$^2$. These studies offer unique information on the strong QCD dynamics that
govern the generation of $N^*$ states with different quantum numbers and distinctively different structural features. The description of the 
structure of the ground state nucleon and the $N(1440)1/2^+$ resonance within the continuum Quantum Chromodynamics (QCD) approach with a traceable 
connection to QCD has demonstrated distinctive differences in their parton distribution amplitudes (PDAs)~\cite{Mezrag:2017znp}. Studies of the
resonant contributions to inclusive electron scattering with the resonance electroexcitation amplitudes from the CLAS data have revealed pronounced 
differences in their evolution with $Q^2$ for the invariant mass $W$ of the final state hadrons within the first, second, and third resonance regions
\cite{HillerBlin:2019hhz}. Therefore, the electroexcitation amplitudes of all prominent resonances in a wide range of $Q^2$ are of particular 
importance in order to explore how $N^*$ states of different structure emerge from QCD. 

\begin{figure*}
  \includegraphics[width=0.55\textwidth]{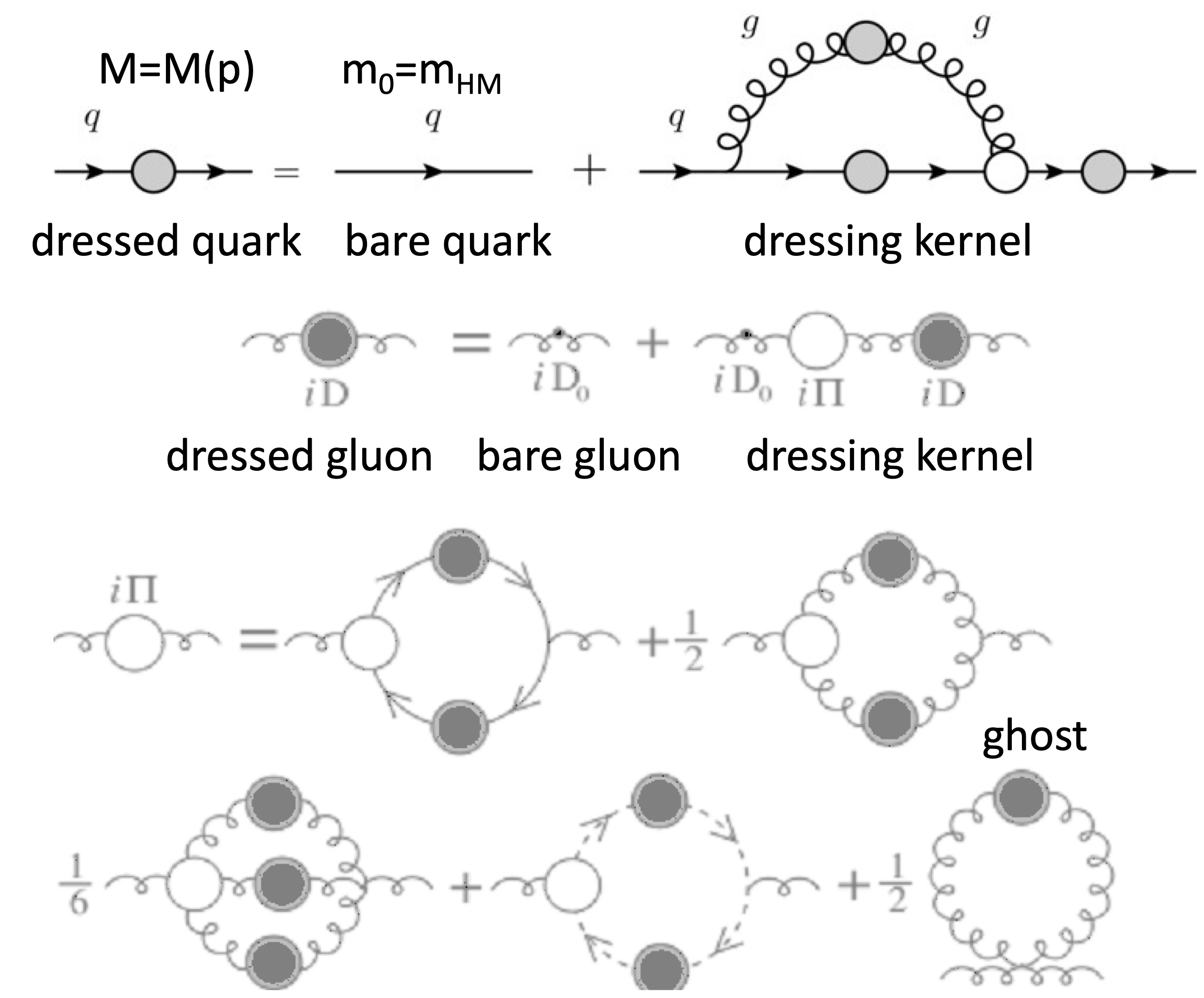}
  \raisebox{9mm}{\includegraphics[width=0.43\textwidth]{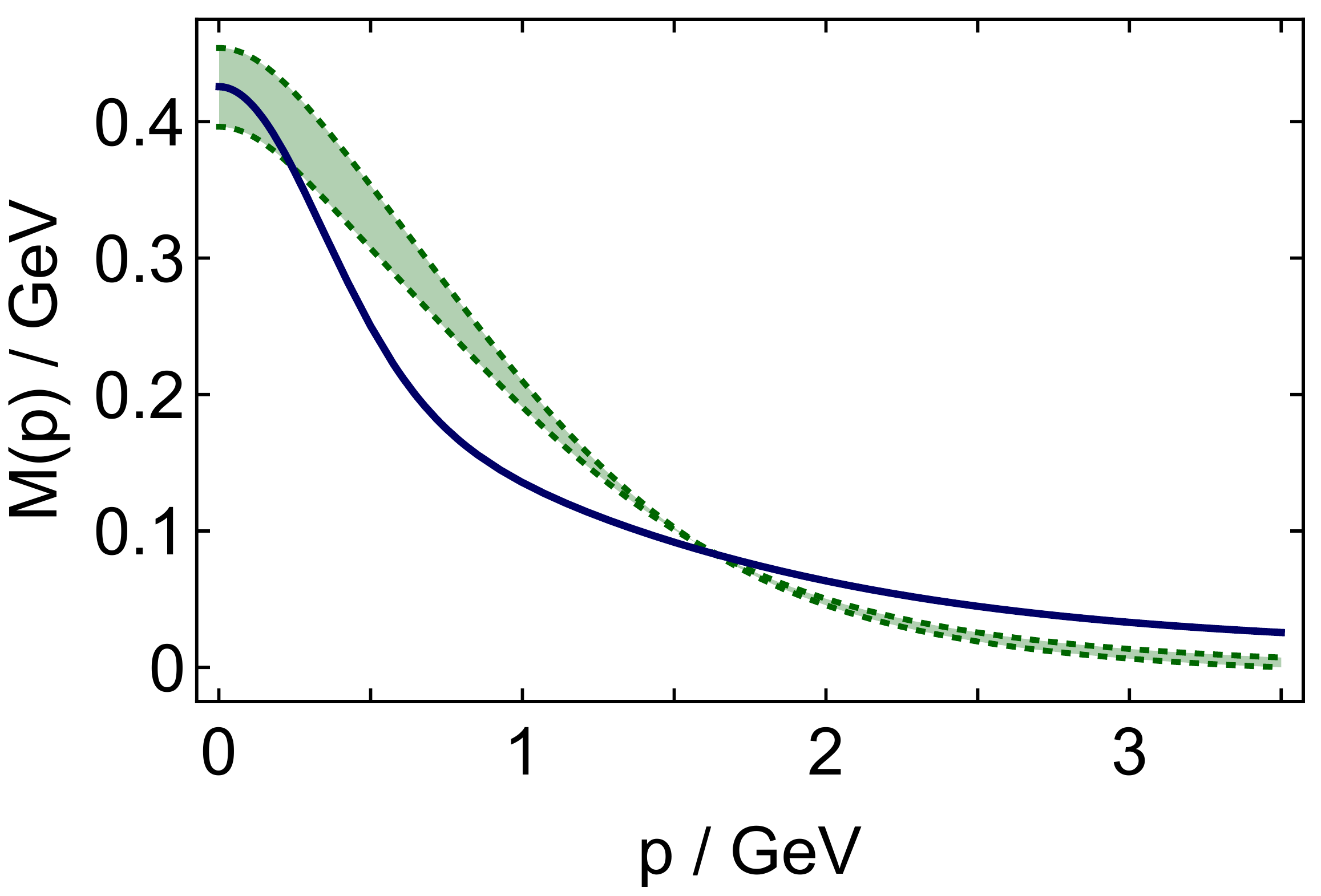}}
\caption{(Left) The mechanisms responsible for the generation of dressed quarks and gluons from bare QCD quarks and gluons within the continuum 
QCD approach~\cite{Roberts:2015lja}. (Right) The momentum dependence of the dressed quark mass (green area) computed from the QCD Lagrangian 
compared with the fit from a phenomenological parameterization to the data on hadron spectra~\cite{Chen:2017pse}.}
\label{mass_dse}
\end{figure*}

Continuum QCD studies have demonstrated that the processes that are responsible for the generation of the dominant part of hadron mass have a 
profound impact on the $Q^2$-evolution of the $\gamma_vpN^*$ electrocouplings~\cite{Roberts:2018hpf}. The bare QCD quarks and gauge gluons 
are affected by the processes imposed by the QCD Lagrangian shown in Fig.~\ref{mass_dse} (left)~\cite{Roberts:2015lja}, which lead to dynamically 
generated dressed quarks and gluons with momentum-dependent masses. The momentum dependence of the dressed quark mass for the light $u$ and $d$ 
quarks computed within continuum QCD is shown in Fig.~\ref{mass_dse} (right)~\cite{Chen:2017pse}. The dominant part of the dressed quark mass and, 
consequently, of the masses of hadrons consisting of light quarks, is generated in the regime of large QCD running coupling ($\alpha_s > 0.3$), 
where the dynamical quark mass increases sharply as the quark momentum decreases, producing a fully dressed light quark of mass 350-400~MeV from 
almost massless (few MeV) bare quarks. The dressing processes shown in Fig.~\ref{mass_dse} (left) are responsible for dynamical QCD chiral symmetry 
breaking (DCSB) on the hadron scale, and represent the major mechanism for the generation of hadron mass in the universe. 

Gaining insight into the dressed quark mass function represents one of the most challenging tasks for experimental hadron physics. Quark-gluon 
confinement prevents the existence of free quarks and gluons, making direct measurements of the dressed quark mass impossible. In $N^*$ 
electroexcitation, the virtual photon interaction with a dressed quark is sensitive to the quark propagator and, therefore, to the running quark mass. 
Consistent results on the momentum dependence of the dressed quark mass from independent studies of the pion and nucleon elastic form factors
\cite{Aguilar:2019teb,Segovia:2014aza} and from the results on the $Q^2$-evolution of the electrocouplings of different resonances
\cite{Segovia:2016zyc,Chen:2018nsg} validates the relevance of dressed quarks with a dynamically generated mass as the main structural components 
within the quark core of hadrons and demonstrates access to the dressed quark mass function in a nearly model-independent way
\cite{Segovia:2015hra,Mokeev:2015lda}. These findings make the studies of the electrocouplings, together with the exploration of the ground state 
nucleon and the structure of the pion, a central emphasis in contemporary hadron physics.

\section{Formalism of Exclusive Resonance Electroproduction}
\label{generalities}

Nucleon resonance electroexcitation can be fully described in terms of three resonance electroexcitation amplitudes or $\gamma_vpN^*$ electrocouplings.
Two of them, $A_{1/2}(Q^2)$ and $A_{3/2}(Q^2)$, describe resonance production in the process $\gamma_v p \to N^*, \Delta^*$ by transversely polarized 
photons of helicity $+1$ in the center-of-mass (CM) frame with the resonance spin projection directed along the $\gamma_v$ momentum equal to 1/2 and 
3/2, respectively. $S_{1/2}(Q^2)$ describes resonance electroexcitation by a longitudinal virtual photon of zero helicity with the resonance spin 
projection equal to 1/2. These electrocouplings are unambiguously determined through their relation with their electromagnetic decay width to the final 
state with transversely $\Gamma_\gamma^T$ and longitudinally $\Gamma_\gamma^L$ polarized photons as: 
\begin{eqnarray}
\Gamma_\gamma^T(W=M_r,Q^2)=\frac{q^2_{\gamma,r}(Q^2)}{\pi}\frac{2M_N}{(2J_r+1)M_r}\times\left(|A_{1/2}(Q^2)|^2+|A_{3/2}(Q^2)|^2\right),\nonumber\\
\Gamma_\gamma^L(W=M_r,Q^2)=\frac{q^2_{\gamma,r}(Q^2)}{\pi}\frac{2M_N}{(2J_r+1)M_r}|S_{1/2}(Q^2)|^2,\label{Eq:EMWidths}
\end{eqnarray}
\noindent
with $q_{\gamma,r}=\left.q_{\gamma} \right|_{W=M_r}$ the absolute value of the $\gamma_v$ three momentum at the resonance point, $M_r$ and $J_r$ 
the resonance mass and spin, respectively, and $M_N$ the nucleon mass. $W$ is the sum of the energies of the $\gamma_v$ and target proton in their 
CM frame. This definition of the $\gamma_vpN^*$ electrocouplings allows them to be related to the helicity amplitudes for resonance electroexcitation 
in any reaction model by making an identity between the resonance electromagnetic decay widths computed within the model and Eq.~(\ref{Eq:EMWidths}). 
In the same way, theory results on the resonance electroexcitation amplitudes can be converted into the electrocouplings.

\begin{figure*}
  \includegraphics[width=0.95\textwidth,center]{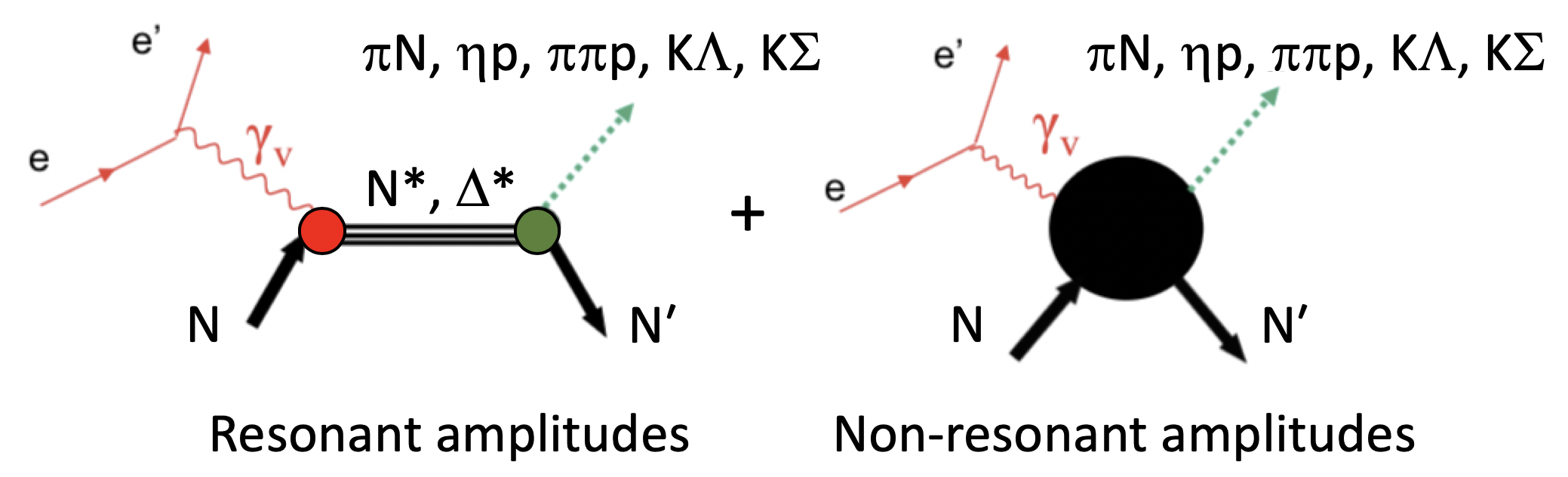}
  \vspace{-5mm}
\caption{Resonant and non-resonant amplitudes of exclusive meson electroproduction channels in the resonance region.}
\label{exclusive_reactions1}
\end{figure*}

Alternatively, the resonance electroexcitation can be described by three transition form factors $G_{1,2,3}(Q^2)$ or $G^*_{M,E,C}(Q^2)$. Two 
of them are relevant for resonances of spin-1/2~\cite{Aznauryan:2011qj}. In the latter case, the $F^*_{1,2}(Q^2)$ Dirac and Pauli transition form 
factors can also be used~\cite{Obukhovsky:2019xrs}, which represent Lorentz invariant functions in the most general expressions for the $N \to N^*$ 
transition electromagnetic currents. The description of the resonance electroexcitation in terms of the electrocouplings and the transition 
electromagnetic form factors is completely equivalent, since they are unambiguously related as described in
Refs.~\cite{Aznauryan:2011qj,Obukhovsky:2019xrs}.

The resonance electrocouplings have been obtained from fits of the observables for several exclusive meson electroproduction channels within specific 
reaction models (see Section~\ref{phenom}). The available observables include differential cross sections, as well as beam, target, and beam-target 
polarization asymmetries. The full amplitude of any meson electroproduction channel is given by the sum of the contributions from all resonances 
excited in the $\gamma_v p$ $s$-channel with hadronic decays to the particular final state and from the complex set of the non-resonant (background) 
contributions (see Fig.~\ref{exclusive_reactions1}). Overall, there are six independent complex amplitudes for both single and double pseudoscalar 
meson and baryon final states. Data analyses within reaction models allow us to isolate the resonant contributions in the full reaction amplitudes. 
In turn, the resonant contributions are related to the electrocouplings that become available at the resonant point $W=M_r$on the real energy axis. 
The first results on the $\Delta(1232)3/2^+$ and $N(1440)1/2^+$ electrocouplings were obtained in a global photo-/hadro-/electroproduction analysis 
of eight final hadron states within the advanced coupled-channel approach developed by the Argonne-Osaka group~\cite{Kamano:2018sfb}. In this analysis 
the resonance parameters were determined from the residues at the pole positions in the complex energy plane.

\section{Studies of Nucleon Resonances in Exclusive Electroproduction}
\label{sec:1}

Studies of nucleon resonances in different exclusive meson electroproduction channels are of particular importance for the extraction of the
electrocouplings, as there are substantial differences in the non-resonant contributions in these different channels. However, the electrocouplings 
should be the same, since the resonance electroexcitation and hadronic decay amplitudes are independent. Therefore, consistent results on the 
electrocouplings from independent studies of different exclusive meson electroproduction channels is of particular importance in order to validate 
their reliable extraction with minimal model dependence. 

Moreover, a successful description of the electroproduction data with the same, $Q^2$-independent, resonance masses, along with their total and 
partial hadronic decay widths in a wide range of $Q^2$, validates the resonance contributions. Several new long-awaited ``missing" resonances have 
been discovered in analyses of exclusive meson photoproduction data~\cite{Bur17}. In addition, the new $N'(1720)3/2^+$ resonance has recently been 
discovered in the combined studies of CLAS $\pi^+\pi^-p$ photo- and electroproduction data~\cite{Mok20}. Future combined studies of meson 
photo-/electroproduction data will extend the available knowledge on the spectrum and structure of nucleon resonances.

Systematic studies of resonance electroexcitation became feasible only after the extensive measurements with CLAS became available for several different
electroproduction channels~\cite{Aznauryan:2011qj,Mokeev:2018zxt,Mokeev:2019ron,Burkert:2019opk,Burkert:2018oyl}. Studies of $N^*$ states will be 
extended with the new CLAS12 detector in the 12~GeV era of experiments at JLab in the range $Q^2 > 5~$GeV$^2$, the highest ever achieved in the study
of exclusive reactions.

\subsection{The CLAS and CLAS12 Detectors in Hall~B}
\label{clas}

The study of the spectrum and structure of excited nucleon states with the CEBAF Large Acceptance Spectrometer (CLAS) in Hall~B~\cite{mecking}, 
referred to as the $N^*$ program, represents one of the experimental physics cornerstones at JLab. In the period from 1997 to 2012 this detector was 
used for studies of inclusive, semi-inclusive, and exclusive reactions from a fixed target with beams of electrons and photons at energies up to 6~GeV.
The efforts of the CLAS Collaboration served to deliver on the science program, posing and sharpening questions that are central to our understanding 
of strong QCD.

CLAS allowed for the study of exclusive reactions in the range of $Q^2$ up to 5~GeV$^2$ and $W$ up to 3~GeV. These data have provided has provided the
dominant part of the available world information on the $\pi N$, $\eta p$, $K \Lambda$, $K \Sigma$, and $\pi^+\pi^- p$ electroproduction channels in the 
resonance region ($W < 2.6$~GeV) with almost complete coverage of the final state CM phase space. The majority of the experiments were carried out with 
an unpolarized liquid-hydrogen target and a longitudinally polarized electron beam.

The large acceptance of this detector is of particular importance for the extraction of the electrocouplings
\cite{Mokeev:2018zxt,Mokeev:2019ron,Burkert:2019opk,Burkert:2018oyl}. Around 200k data points for differential cross sections, separated 
structure functions, and single- and double-polarization observables have become available based on analyses of these experimental data, which 
are stored in the CLAS Physics Database~\cite{db}. These measurement data points continue to increase as additional analyses are completed.

In the period from 2012 to 2017 CLAS was replaced with the new large acceptance CLAS12 spectrometer~\cite{clas12-nim} as part of the JLab 12~GeV 
upgrade project. The extended program includes a number of experiments as part of the continuing $N^*$ program in Hall~B, which will collect data 
over an unprecedented kinematic range for the study of nucleon excited states in the range of $Q^2$ from 0.05 to at least 9~GeV$^2$, spanning the 
full CM angular range of the decay final states.

\begin{figure*}
\centering
  \includegraphics[width=0.9\textwidth]{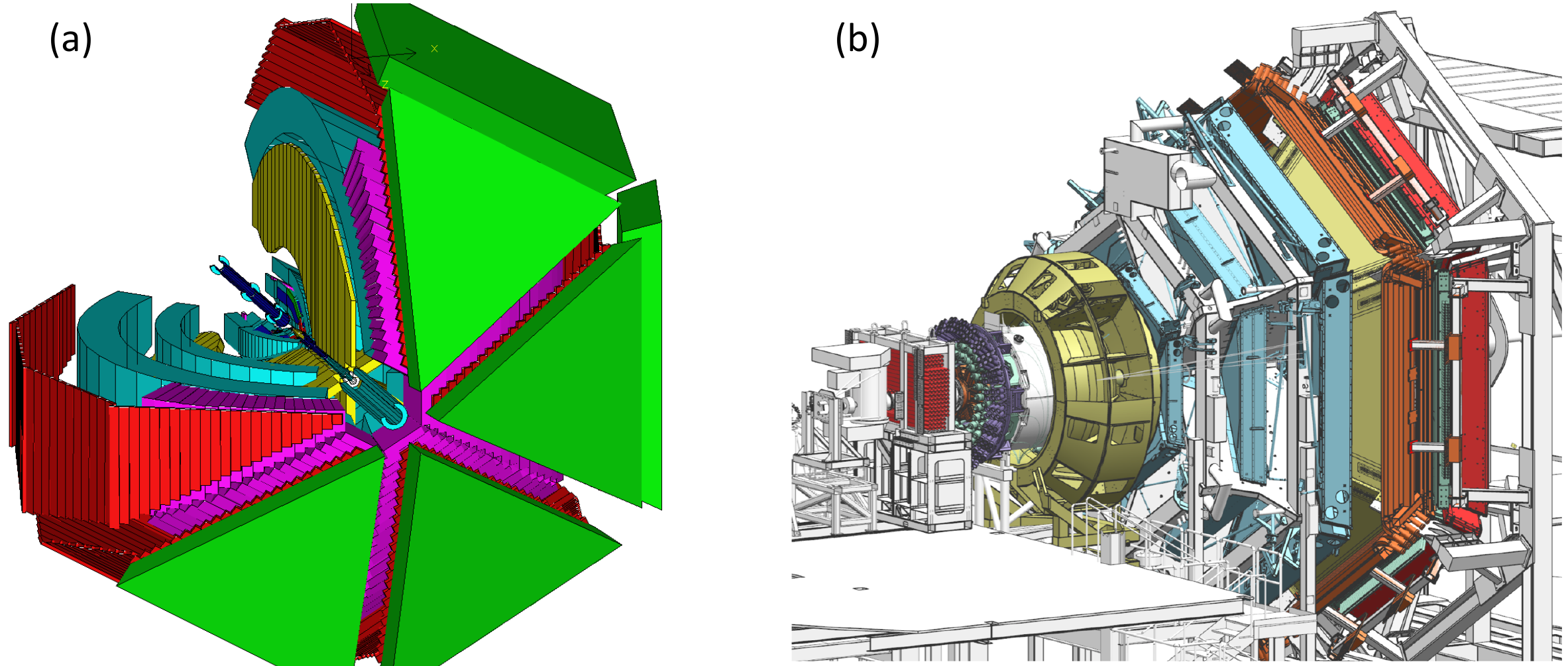}
\caption{(a) Model of the original CLAS spectrometer for use with electron and photon beams up to energies of 6~GeV with one of the six sectors 
removed to enable a view within the detector. The overall length of the detector is $\sim$6~m. (b) Model of the new CLAS12 spectrometer for use 
with electron beams up to 11~GeV. The overall length of the detector is $\sim$15~m.}
\label{clas-versions}
\end{figure*}

Figure~\ref{clas-versions} shows model pictures of the original and the new spectrometers. The CLAS detector consisted of six equivalent azimuthal 
sectors around a superconducting torus magnet that spanned polar angles from 5$^\circ$ to 140$^\circ$ for beam energies up to 6~GeV. It was designed 
to operate with photon beams up to $1 \times 10^7$~$\gamma$/s and with electron beams at a luminosity of 1$\times$10$^{34}$~cm$^{-2}$s$^{-1}$. The new 
CLAS12 spectrometer is composed of a Forward Detector system built around a superconducting torus magnet that spans polar angles from 5$^\circ$ to 
35$^\circ$ and a Central Detector built around a superconducting solenoid magnet that spans polar angles from 35$^\circ$ to 125$^\circ$. This device 
has been optimized for the reconstruction of exclusive reactions at beam energies up to 11~GeV. Both detectors were designed to use drift chambers for 
charged particle tracking, time-of-flight hodoscopes for precise timing measurements for particle identification, and a forward electromagnetic 
calorimeter for electron and neutral identification. In addition, CLAS12 includes a central silicon+Micromegas tracker, specialized Cherenkov systems 
for electron and hadron identification, and a compact forward electron tagger for quasi-real photoproduction measurements at $Q^2$ as low as 
0.05~GeV$^2$. The nominal operating luminosity of CLAS12 is 1$\times$10$^{35}$~cm$^{-2}$s$^{-1}$ in order to allow measurements at the highest 
possible $Q^2$.

\subsection{Progress in Experimental Studies of $\pi N$ Electroproduction}
\label{npi_channels}

Studies of exclusive $\pi^0p$ and $\pi^+n$ electroproduction represent an effective tool for the exploration of the structure of excited nucleon 
states. Resonances in the mass range $W < 1.6$~GeV decay preferentially to $\pi N$. For these states, the studies of $\pi N$ electroproduction 
are the driving force in the exploration of their structure. Several resonances in the mass range $W > 1.6$~GeV still have appreciable branching 
fractions for decays to $\pi N$. Information on the electrocouplings of these states from $\pi N$ electroproduction can be compared with the 
results from other exclusive channels, allowing for cross checks of their systematic uncertainties. 

\begin{figure*}
\centering
\includegraphics[width=0.8\textwidth]{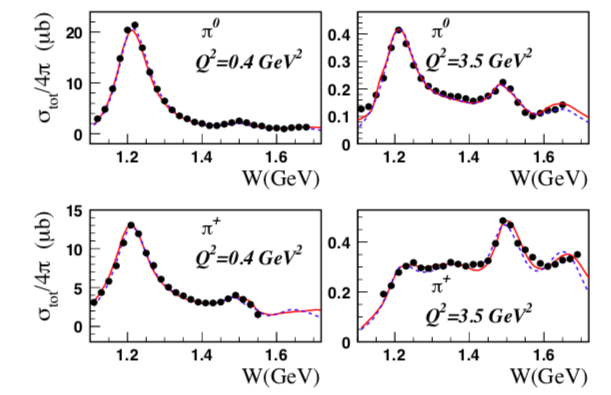}
\vspace{-2mm}
\caption{$\pi^0p$ (top) and $\pi^+n$ (bottom) fully integrated cross sections from CLAS~\cite{Joo:2001tw,Egiyan:2006ks,Park:2007tn} 
and their description within dispersion relation (solid red line) and unitary isobar (blue dashed line)~\cite{Aznauryan:2002gd} models.}
\label{npi_integ}
\end{figure*}

The CLAS detector has provided the dominant part of the available world information on exclusive $\pi N$ electroproduction. A total of nearly 
160k data points on unpolarized differential cross sections, longitudinally polarized beam asymmetries, and longitudinally polarized target 
and beam-target asymmetries have become available with almost complete coverage of the final state phase space
\cite{Aznauryan:2011qj,Park:2014yea,Markov:2019fj,Joo:2001tw,Egiyan:2006ks,Park:2007tn,addk1,addk2,addk3,addk4,addk5}. The kinematic coverage 
of the $\pi N$ data is presented in Table~\ref{1pi-coverage}. Representative examples of the $W$ dependence of the fully integrated $\pi^0p$ and 
$\pi^+n$ cross sections~\cite{Joo:2001tw,Egiyan:2006ks,Park:2007tn} are shown in Fig.~\ref{npi_integ}. 

\begin{table*}[htb]
\begin{center}
\vspace{2mm}
\begin{tabular}{|c|c|c|c|} \hline
$\pi N$ Final  & $W$ Coverage,  & $Q^2$ Coverage &  Measured      \\
State          & GeV            & GeV$^2$       &  Observables   \\ \hline
               &  1.1-1.38      & 0.16-0.36      & $d\sigma/d\Omega$ \\
 $\pi^+n$      &  1.1-1.55      & 0.30-0.60      & $d\sigma/d\Omega$  \\ 
               &  1.1-1.70      & 1.7-4.5        & $d\sigma/d\Omega$, $A_b$ \\  
               &  1.6-2.00      & 1.7-4.5        & $d\sigma/d\Omega$ \\ \hline
               &  1.1-1.38      & 0.16-0.36      & $d\sigma/d\Omega$ \\  
 $\pi^0p$      &  1.1-1.68      & 0.40-1.80      & $d\sigma/d\Omega$, $A_b$, $A_t$, $A_{bt}$ \\ 
               &  1.1-1.39      & 3.00-6.00      & $d\sigma/d\Omega$ \\
               &  1.4-1.90      & 0.40-1.00      & $d\sigma/d\Omega$ \\ \hline               
\end{tabular}
\end{center}
\caption{$Q^2$ and $W$ ranges covered by the CLAS $\pi N$ electroproduction data and the measured observables: differential virtual photon cross 
sections $d\sigma/d\Omega$, and longitudinally polarized beam, target, and beam-target asymmetries, $A_b$, $A_t$, and $A_{bt}$, respectively.}
\label{1pi-coverage}
\end{table*}

The resonance structures in the first and second resonance regions demonstrate pronounced and different dependencies in both channels as a
function of $Q^2$. At $Q^2 \sim 0.4$~GeV$^2$, the $\Delta(1232)3/2^+$ resonance represents the most pronounced structure. Instead, at 
$Q^2 \sim 3.5$~GeV$^2$, the second resonance region becomes the most pronounced structure in the $\pi^+n$ channel, comparable with the 
$\Delta(1232)3/2^+$ in the $\pi^0p$ channel. These features emphasize the sensitivity of both the $\pi^0p$ and $\pi^+n$ channels to the 
resonant contributions. Pronounced differences in the structure of the resonances located in the first and second resonance regions are
responsible for the $Q^2$ evolution of the $\pi N$ integrated cross sections.

\begin{figure*}
\centering
\includegraphics[width=0.95\textwidth]{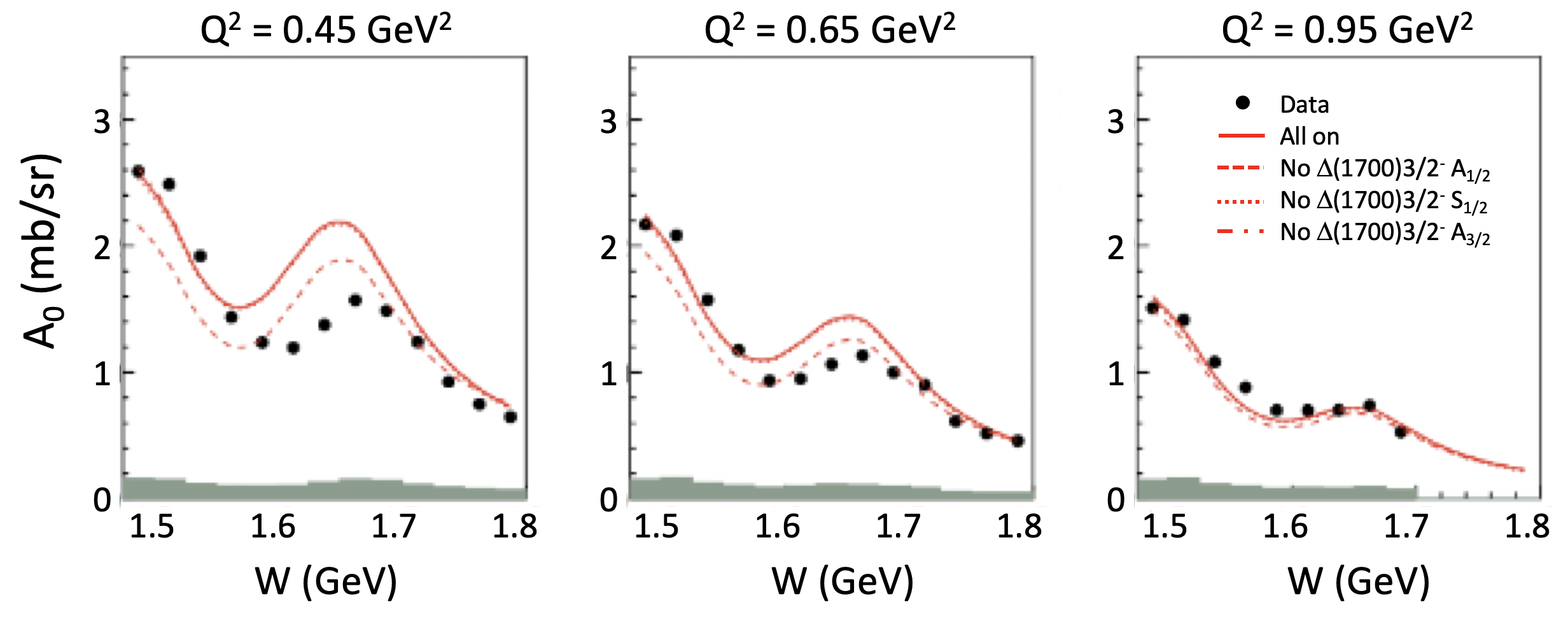}
\vspace{-2mm}
\caption{$A_0$ Legendre moments at different $Q^2$ as a function of $W$ in comparison with model calculations~\cite{Aznauryan:2002gd} for the 
electrocouplings of the resonances from the CLAS data~\cite{Mokeev:2018zxt,Mokeev:2015lda} showing sensitivity to the $\Delta(1700)3/2^-$ by 
turning on/off the resonance electrocouplings. For the definition of the Legendre moments see Ref.~\cite{Aznauryan:2011qj}.}
\label{delta12_delta_32}
\end{figure*}

The isospin invariance through the Clebsch-Gordon coefficients makes the $\pi^0p$ channel particularly sensitive to the contributions from 
$\Delta^*$ resonances of isospin 3/2. Precise data on differential cross sections and polarization asymmetries in the first resonance region 
for the $\pi^0p$ channel has provided information on the dominant $N \to \Delta(1232)3/2^+$ magnetic transition form factor for 
$0.16 < Q^2 < 6.0$~GeV$^2$~\cite{Aznauryan:2009mx}. These CLAS results provide insight into the dynamics responsible for hadron mass 
generation, as described in Section~\ref{EHM}. Furthermore, from the studies of $\pi^0p$ and $\pi^+n$ electroproduction in the first resonance 
region, information on the $R_{EM}$ and $R_{SM}$ ratios of the electric quadrupole and scalar quadrupole amplitudes relative to the leading magnetic 
dipole amplitude, respectively, have become available~\cite{Aznauryan:2011qj,Aznauryan:2009mx}. At small $Q^2$, the non-zero values of $R_{EM}$ 
and $R_{SM}$ offer evidence for the departure from a spherical shape of the $\Delta(1232)3/2^+$. The behavior of these ratios at $Q^2$ up to 
6~GeV$^2$ demonstrates that at this distance scale, the electroexcitation of the $\Delta(1232)3/2^+$ remains inconsistent with expectations from 
perturbative QCD (pQCD)~\cite{Aznauryan:2011qj}.

Precise data on the $\pi^0p$ and $\pi^+n$ differential cross sections, the unpolarized, transverse-transverse, and longitudinal-transverse 
structure functions, as well as the results on the longitudinally polarized beam, target, and beam-target asymmetries have become available 
from CLAS in the second resonance region of $1.4< W < 1.6$~GeV at $0.2 < Q^2 < 5.0$~GeV$^2$. Analyses of these experimental data have provided 
precise information on the electroexcitation amplitudes of the $N(1440)1/2^+$, $N(1520)3/2^-$, and $N(1535)1/2^-$ for $Q^2 < 5$~GeV$^2$ for the 
first time~\cite{Aznauryan:2009mx}. 

Studies of $\pi N$ electroproduction continue to gradually extend over $W$, allowing us to explore the electroexcitation amplitudes of the 
states in the third resonance region. Studies of $\pi^+n$ electroproduction for $1.5 < Q^2 < 5.0$~GeV$^2$~\cite{Park:2014yea} have provided 
information on the electrocouplings of the $N(1675)5/2^-$ and $N(1710)1/2^+$ for the first time and extended the results on the electrocouplings 
of the $N(1680)5/2^+$ to $Q^2 > 1.5$~GeV$^2$. These resonances decay preferentially to the $\pi N$ final states and there are no close resonances 
of isospin 3/2 with the same spin and parity. Hence, analysis of only $\pi^+n$ electroproduction is sufficient to extract their electrocouplings.

Recently published CLAS results on $\pi^0p$ electroproduction for $1.4 < W < 1.9$~GeV and $0.4 < Q^2 < 1.0$~GeV$^2$~\cite{Markov:2019fj} make it 
possible to explore $\Delta^*$ states in the third resonance region. The substantial sensitivity of these new data to the electrocouplings of the
$\Delta(1700)3/2^-$ is highlighted in Fig.~\ref{delta12_delta_32}. Here the $A_0$ Legendre moments from the $\pi^0p$ cross sections are compared 
with model expectations~\cite{Aznauryan:2002gd} when the electrocouplings of this $\Delta^*$ state available from the $\pi^+\pi^-p$ data
\cite{Mokeev:2018zxt,Mokeev:2015lda} are turned on/off.

The CLAS $\pi^+n$ data should be extended within the range $1.4 < W < 1.9$~GeV and $0.4 < Q^2 < 1.0$~GeV$^2$ covered by the $\pi^0p$ data 
\cite{Markov:2019fj}. Also, data from the $\pi^0p$ channel should be made available for $1.4 < W < 2.0$~GeV and $1.5 < Q^2 < 5.0$~GeV$^2$ covered 
by the existing $\pi^+n$ data~\cite{Aznauryan:2011qj,Park:2014yea}. The combined analysis of the $\pi^+n$ and $\pi^0p$ data in the same range of 
$W$ and $Q^2$ within the available models~\cite{Aznauryan:2002gd} represents an important part of the efforts aimed at obtaining the electrocouplings 
of the states in the third resonance region with isospin $I = 1/2$ and 3/2. These data also provide important input for the development of 
coupled-channel analyses.

\subsection{$\pi^+\pi^-p$ and $KY$ Exclusive Electroproduction in the Resonance Region}
\label{kypipip}

The $\gamma_v p \to \pi^+\pi^-p$ electroproduction channel is sensitive to the most excited states observed so far in the $N^*$ spectrum. $N^*$ states 
in the mass range $<$1.6~GeV decay mostly to $\pi N$ final states~\cite{Tanabashi:2018oca}, making the $\pi N$ electroproduction channels the primary 
source of information on their electrocouplings. However, these low-lying resonances, except for the $N(1535)1/2^-$, also have significant branching 
fractions (above 30\%) for decays into the $\pi \pi N$ final states, which also makes it possible to evaluate their electrocouplings through 
$\pi^+\pi^-p$ photo-/electroproduction in this channel. Consistent results on the electrocouplings of nucleon resonances determined from independent 
studies of the $\pi N$ and $\pi^+\pi^-p$ channels allow for cross checks of systematic uncertainties, in particular those associated with the reaction 
models.

Studies of $\pi^+\pi^-p$ electroproduction are of particular importance in order to gain insight into the electrocouplings of $N^*$ states in the 
third resonance region. Here, the $\Delta(1620)1/2^-$, $\Delta(1700)3/2^-$, and $N(1720)3/2^+$ have branching fractions for decays of $>$50\%. For
this final state, the branching fractions for many resonances in the third resonance region have been established with large uncertainties
\cite{Tanabashi:2018oca}. However, almost all resonances in the mass range from 1.6~GeV to 1.75~GeV, except for the $N(1675)5/2^-$ and $N(1710)1/2^+$, 
have a measurable impact on the $\pi^+\pi^-p$ observables, making studies of this exclusive channel an important tool for insight into the 
electroexcitation amplitudes of most states in the third resonance region. Furthermore, a successful description of $\pi^+\pi^-p$ electroproduction data 
with $Q^2$-independent resonance hadronic decay widths allows us to improve the knowledge on hadronic decays of these states~\cite{Mokeev:2015lda,Mok20}.  

According to the quark model results on the hadronic decays of $N^*$ states~\cite{Capstick,Santopinto}, studies of $\pi^+\pi^-p$ 
photo-/electroproduction offer promising opportunities in the search for the missing resonances. The existence of these states is predicted based 
on the approximate SU(6) symmetry of the strong interaction relevant in the regime of strong QCD. Searches for these states for a long time have been 
a challenging task. Recently, a new $N'(1720)3/2^+$ resonance was observed in the combined studies of $\pi^+\pi^-p$ photo- and electroproduction data
\cite{Mok20}, in addition to several new missing resonances discovered in a global multi-channel analysis of meson photo- and hadroproduction
\cite{Bur17}. For the first time the results on the electrocouplings of the missing $N'(1720)3/2^+$ resonance have become available, offering insight 
into its internal structure. 

\begin{figure*}
\centering
  \includegraphics[width=0.7\textwidth]{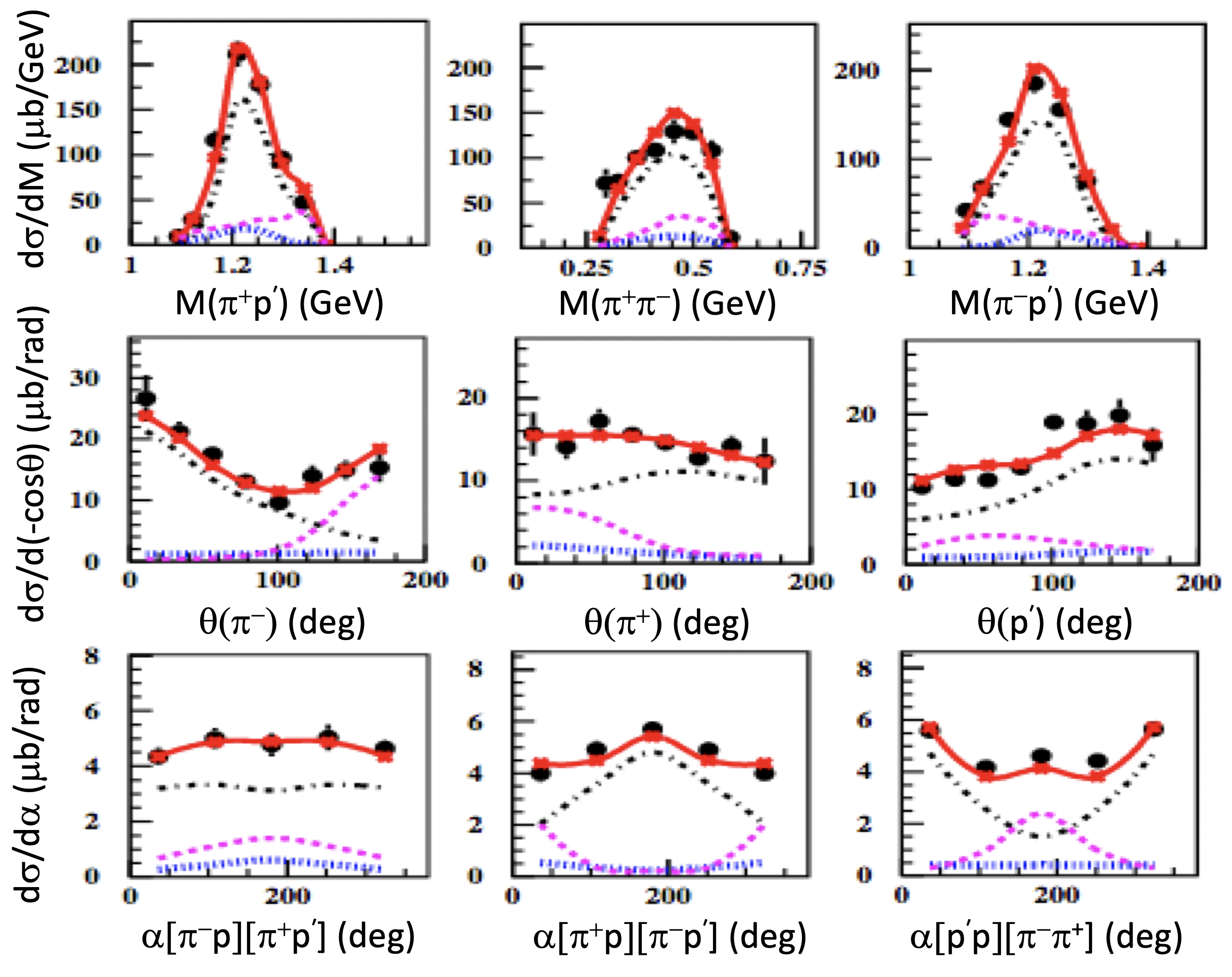}
  \includegraphics[width=0.7\textwidth]{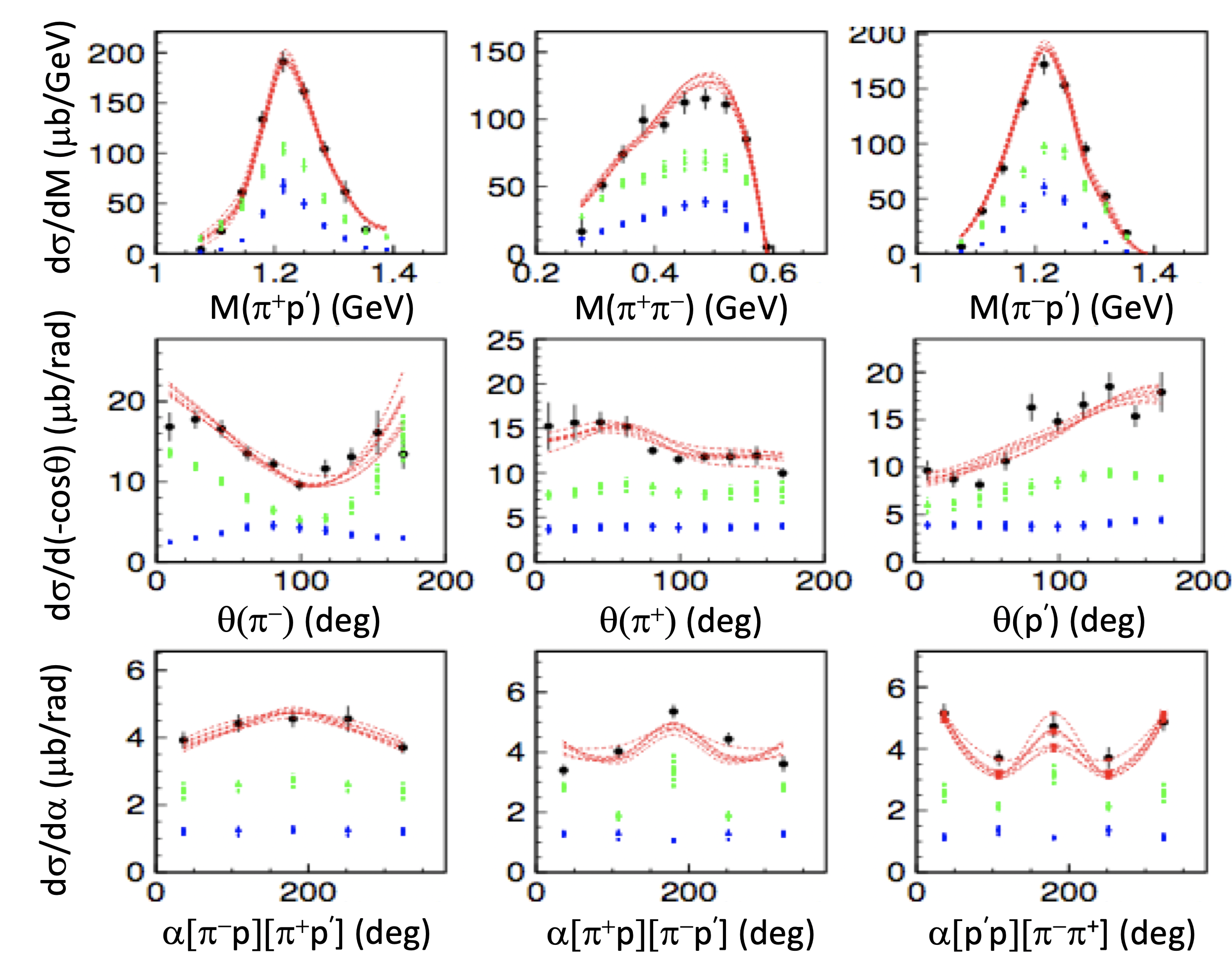}
\caption{Nine 1-fold differential $\pi^+\pi^-p$ electroproduction cross sections measured with CLAS at $W$=1.51~GeV and $Q^2$= 0.65~GeV$^2$
\cite{Ripani:2002ss}. (Top) Data description~\cite{Mokeev:2015lda} within the JM model outlined in Section~\ref{phenom_2pi} (red solid lines) and 
the contributing meson-baryon channels with amplitudes established from the data: $\pi^-\Delta^{++}$ (black dot-dashed), $\pi^+\Delta^0$ (blue 
dashed), 2$\pi$ direct production (magenta dashed). (Bottom) Fit of the data within the JM model. Selected cross sections computed in the data fit 
are shown by the red dashed curves. The resonant/non-resonant contributions determined in the data fit are shown by the blue/green bars.}
\label{exclusive_reactions2}
\end{figure*}

The CLAS detector has provided the only available world data on the nine independent, unpolarized 1-fold differential $\pi^+\pi^-p$ photo- and 
electroproduction cross sections in the broad kinematic range of $W$ and $Q^2$ listed in Table~\ref{pipip_kin_areas}. Representative examples 
of these differential cross sections in a particular bin of $W$ and $Q^2$ are shown in Fig.~\ref{exclusive_reactions2}. They consist of invariant 
mass distributions for the three pairs of final state hadrons, three CM angular distributions for the final state hadrons, and three CM distributions
for the angles between the reaction plane for one of the final state hadrons with the target proton and a second plane defined by the three momenta 
of the other two final state hadrons. The kinematics of the $\pi^+\pi^-p$ state is fully determined by 5 independent variables. The nine 1-fold 
differential cross sections represent the integrals from the common 5-fold differential cross sections over different sets of 4 variables. In recent
studies~\cite{Trivedi:2018rgo}, nine transverse-transverse and transverse-longitudinal contributions into the $\pi^+\pi^-p$ electroproduction cross 
sections have become available along with the nine unpolarized cross sections.

The CLAS data on $\pi^+\pi^-p$ electroproduction have provided the first results on the electrocouplings of the $N(1440)1/2^+$ and $N(1520)3/2^-$ 
at $Q^2 < 0.55$~GeV$^2$~\cite{Fedotov:2008aa}. Analysis of other  $\pi^+\pi^-p$ data~\cite{Ripani:2002ss} have extended the information on these 
electrocouplings in the $Q^2$ range up to 1.5~GeV$^2$ and provided results on the electrocouplings of the $\Delta(1620)1/2^-$~\cite{Mokeev:2015lda}, 
as well as preliminary results on the electrocouplings of the $N(1680)5/2^+$, $\Delta(1700)3/2^-$, and $N(1720)3/2^+$
\cite{Mokeev:2018zxt,Mokeev:2019ron,Ripani:2002ss}. Analysis of the highest precision $\pi^+\pi^-p$  data~\cite{Fedotov:2018oan}, collected in $Q^2$ 
bins as small as 0.05~GeV$^2$ (see Table~\ref{pipip_kin_areas}) will further improve our knowledge of $N^*$ electroexcitations in the third resonance 
region. Analysis of these data~\cite{Isupov:2017lnd,Trivedi:2018rgo} will provide information on the electrocouplings of all resonances in the mass 
range up to 2~GeV that are prominent for $2 < Q^2 < 5$~GeV$^2$ and will shed light on the manifestations of new baryon states, including those seen 
through a global multi-channel analysis of exclusive photoproduction~\cite{Bur17}.

\begin{table*}
\begin{center}
\vspace{2mm}
\begin{tabular}{|c|c|c|c|} \hline
$Q^2$ Coverage,  & $W$ Coverage,  & $W$/$Q^2$ Bin Size  &  Data Status        \\
GeV$^2$          & GeV            & GeV/GeV$^2$      &                     \\ \hline
    0.           &  1.6-2.0       & 0.25/N.A.        & Completed \cite{Golovatch:2018hjk} \\ \hline 
 0.2-0.6         &  1.30-1.57     & 0.025/0.050      & Completed \cite{Fedotov:2008aa} \\ \hline
 0.5-1.5         &  1.40-2.10     & 0.025/0.3-0.4    & Completed \cite{Ripani:2002ss} \\ \hline 
 0.4-1.0         &  1.30-1.85     & 0.025/0.050      & Completed \cite{Fedotov:2018oan}  \\ \hline
 2.0-5.0         &   1.40-2.00    & 0.025/0.5        & Completed \cite{Isupov:2017lnd} \\ \hline 
 2.0-5.0         &   1.40-2.00    & 0.025/0.5        & In progress \cite{Trivedi:2018rgo}  \\ \hline  
\end{tabular}
\end{center}
\caption{Kinematic area in $Q^2$ and $W$ covered by the CLAS measurements of the $\pi^+\pi^-p$ photo-/electroproduction cross sections and the
associated data references.}
\label{pipip_kin_areas}
\end{table*}

The recent advancements in understanding the spectrum and structure of the nucleon excited states have mainly been provided by advanced 
analyses of the exclusive $\pi N$, $\pi \eta$, and $\pi^+ \pi^- p$ reaction channels. However, with the publication of the high quality 
$K^+ \Lambda$ and $K^+ \Sigma^0$ photoproduction data from CLAS~\cite{bradford2006,bradford2007,mccracken2010,dey2010}, the potential and 
importance of the hyperon channels has been demonstrated. In fact, the spectrum of $N^*$ states listed in the recent edition of the Particle 
Data Group (PDG)~\cite{Tanabashi:2018oca} has been radically altered by the  $K^+Y$ photoproduction data, including both differential cross 
sections and polarization observables. Table~\ref{nstar-evol} shows a comparison of the current PDG listings for a dozen $N^*$ and $\Delta^*$ 
states compared to just a decade ago. For most of these states the $K^+Y$ data were of critical importance in the update. The $K^+\Lambda$ and 
$K^+\Sigma^0$ channels are important to consider separately. Although the two ground-state hyperons have the same valence quark content ($uds$), 
they differ in isospin ($I$=0 for $\Lambda$ and $I$=1 for $\Sigma^0$), so that $N^*$ states of $I=1/2$ can decay to $K^+\Lambda$, but $\Delta^*$ 
states cannot. Because both $N^*$ and $\Delta^*$ resonances can couple to the $K^+\Sigma^0$ final state, the hyperon final state selection 
constitutes an isospin filter.

\begin{table}[htbp]
\centering
\begin{tabular}{c|c|c|c|c|c|c} \hline
State               & PDG      & PDG  & $\pi N$ & $K\Lambda$ & $K \Sigma$ & $\gamma N$ \\
$N(mass)J^P$        & pre-2010 & 2018 &         &            &            &            \\ \hline
$N(1710)1/2^+$      & *** & **** & **** & ** & *  & **** \\ \hline
$N(1875)3/2^-$      &     & ***  & **   & *  & *  & **   \\ \hline
$N(1880)1/2^+$      &     & ***  & *    & ** & ** & **   \\ \hline
$N(1895)1/2^-$      &     & **** & *    & ** & ** & **** \\ \hline
$N(1900)3/2^+$      & **  & **** & **   & ** & ** & **** \\ \hline
$N(2000)5/2^+$      & *   & **   & *    &    &    & **   \\ \hline
$N(2060)5/2^-$      &     & ***  & **   & *  & *  & ***  \\ \hline
$N(2100)1/2^+$      & *   & ***  & ***  & *  &    & **   \\ \hline
$N(2120)3/2^-$      &     & ***  & **   & ** & *  & ***  \\ \hline
$\Delta(1600)3/2^+$ & *** & **** & ***  &    &    & **** \\ \hline
$\Delta(1900)1/2^-$ & **  & ***  & ***  &    & ** & ***  \\ \hline
$\Delta(1940)3/2^-$ & *   & **   & **   &    &    & *.   \\ \hline
$\Delta(2200)7/2^-$ & *   & ***  & **   &    & ** & ***  \\ \hline
\end{tabular}
\caption{Evolution of our understanding of the excited $N^*$ and $\Delta^*$ spectrum over the past decade and the available evidence from 
different initial/final states based on the PDG ``*" ratings in the pre-2010 and 2018 listings.  Note also that our knowledge of more than a dozen 
additional states in the most recent PDG listings (not shown here), while remaining unchanged in their overall * rating, have now been 
improved by quantifying their coupling to additional initial/final states compared to the pre-2010 listings. The $KY$ channels have been a 
vital inclusion in this expansion of our understanding.}
\label{nstar-evol}
\end{table}

Beyond the available $KY$ photoproduction data, the $KY$ electroproduction data are by far the most extensive and precise measurements within the 
nucleon resonance region $1.6 < W < 2.6$~GeV~\cite{carman16,carman18}. They are of great value for independent extractions of the resonance 
electrocouplings available from the $\pi N$ and $\pi^+\pi^-p$ electroproduction data at $W > 1.6$~GeV. Measurements have been provided for the 
differential cross sections and separated structure functions $\sigma_T$, $\sigma_L$, $\sigma_U = \sigma_T + \epsilon \sigma_L$, $\sigma_{LT}$, 
$\sigma_{TT}$, and $\sigma_{LT'}$ for $K^+\Lambda$ and $K^+\Sigma^0$~\cite{raue-car,5st,sltp,carman3}, recoil polarization for $K^+\Lambda$~\cite{ipol}, 
and beam-recoil transferred polarization for $K^+\Lambda$ and $K^+\Sigma^0$~\cite{carman1,carman2}. These data span $Q^2$ from 0.5 to 4.5~GeV$^2$, 
$W$ from 1.6 to 3~GeV, and the full CM angular range of the $K^+$. $KY$ exclusive production is sensitive to coupling to higher-lying $N^*$ states 
for $W > 1.6$~GeV. It is precisely in this mass range where our knowledge of the $N^*$ spectrum is the most limited. These data have comparable 
uncertainties as for the available CLAS $\pi^+ \pi^- p$ electroproduction data and can also be used to confirm the signals of new baryon states 
observed in photoproduction in a complementary fashion by checking whether within each bin of $Q^2$, the determined states have the same masses and 
decay widths. Reliable information on the electrocouplings from the $KY$ channels is not yet available due to the lack of a suitable reaction model. 
Further development of reaction models~\cite{Corthals:2007kc,Ronchen:2018ury}, as well as other approaches that will make it possible to employ them 
for the extraction of the resonance electrocouplings from the $KY$ electroproduction data is critically needed in order to make progress with these 
important channels.

\subsection{Approaches for Evaluation of Resonance Electrocouplings from Data}
\label{phenom}

Most of the results on the $\gamma_vpN^*$ electrocouplings have become available from independent analyses of the major exclusive meson
electroproduction channels $\pi^+n$, $\pi^0p$, and $\pi^+\pi^-p$~\cite{Aznauryan:2011qj,Mokeev:2018zxt,Mokeev:2019ron,Tiator:2018pjq,Tiator:2011pw,said}. 
The electrocouplings of the $N(1535)1/2^-$ were also obtained from the data on $\eta p$ electroproduction~\cite{Denizli:2007tq}. The first 
results on the electrocouplings of the $\Delta(1232)3/2^+$ and $N(1440)1/2^+$ at their pole positions were determined in a global 8-channel 
analysis of exclusive meson electro- and hadroproduction data developed by the Argonne-Osaka group~\cite{Kamano:2018sfb}. These breakthrough results 
should be extended in the future for other excited nucleon states. In this Section we consider approaches developed by the CLAS Collaboration for 
the extraction of the electrocouplings from the data on exclusive $\pi N$ and $\pi^+\pi^-p$ electroproduction.

\subsubsection{Electrocoupling Extraction from CLAS $\pi N$ Electroproduction Data}
\label{phenom_1pi}

The extraction of the electrocouplings from the CLAS $\pi^0p$ and $\pi^+n$ electroproduction data was carried out within two different approaches
that used a) a dispersion relation technique (DR) and b) the Unitary Isobar Model (UIM)~\cite{Aznauryan:2002gd,Aznauryan:2009mx}.

\begin{figure*}
\centering
\includegraphics[width=0.90\textwidth]{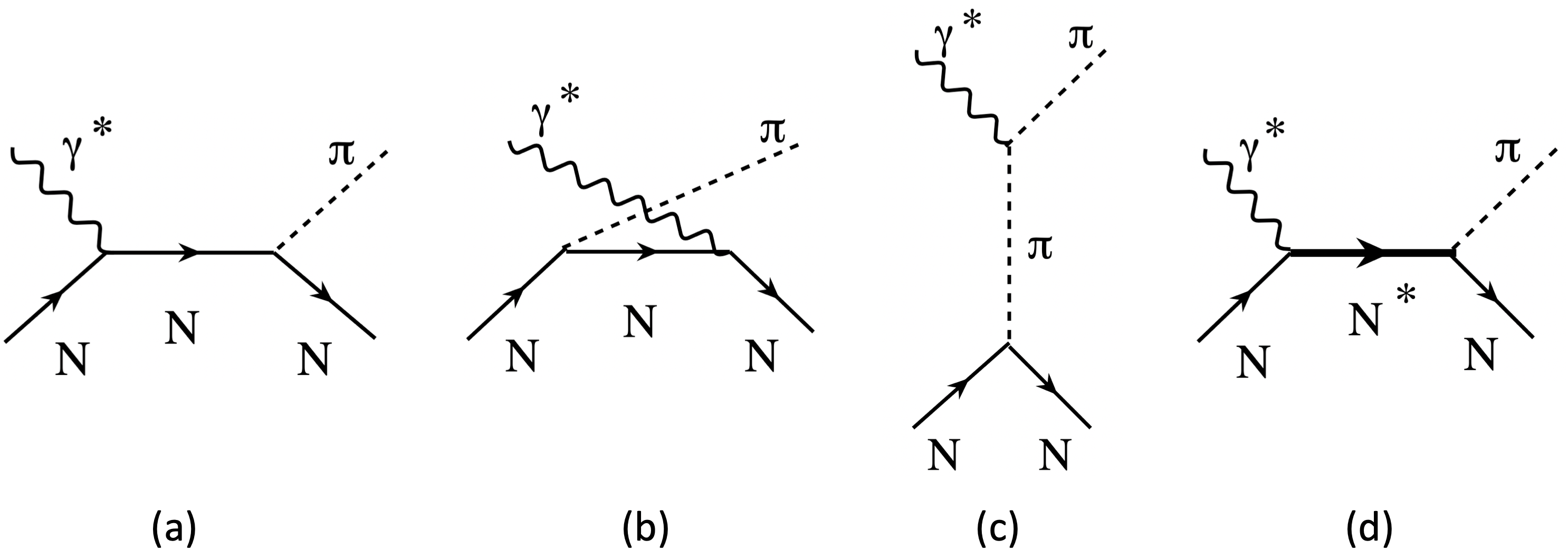}
\caption{Born terms (a, b, c) and the resonant contribution (d) for the description of $\pi^0p$ and $\pi^+n$ electroproduction in
Refs.~\cite{Aznauryan:2011qj,Aznauryan:2002gd,Aznauryan:2009mx}.}
\label{npi_dynamics}
\end{figure*}

In the DR approach, exclusive $\pi N$ electroproduction has been described by six independent Lorentz invariant functions in the most general 
expression for the transition $N \to \pi N$ electromagnetic current. These functions, the so-called Ball amplitudes, are defined in 
Ref.~\cite{Aznauryan:2011qj}. Each Ball amplitude $B_i(s,t)$ ($i$=1,2,3,6,8) and $B'_5(s,t)$ in both the $\pi^0p$ and $\pi^+n$ channels represents 
a complex function of the Mandelstam variables $s$ and $t$. At fixed values of $t$ and $Q^2$ the real part of the amplitudes $Re[B_i(s,t)]$ can be 
expressed through their imaginary parts $Im[B_i(s,t)]$ by employing dispersion relations~\cite{Aznauryan:2011qj,Aznauryan:2002gd,Aznauryan:2009mx}. 
For the $B_3$ amplitudes, what is known as the subtraction procedure was employed in the dispersion relations in order to provide a zero value for 
the dispersive integrand at absolute values of $s$ in the complex energy plane that tend to infinity. This procedure invokes a phenomenological
subtraction function $f_{sub}(t,Q^2)$ fit to the data. For the description of the $\pi N$ reaction from both protons and neutrons, and accounting 
for isospin, we have 15 unsubtracted and 1 subtracted dispersion relations, which are based just on the most general principles of analyticity and 
crossing symmetry for the reaction amplitudes. The amplitude singularities and residues that enter into the dispersion relations were obtained 
assuming pion exchange in the $t$-channel and nucleon exchange in the $s$-channel as shown in Fig.~\ref{npi_dynamics}. Eventually, pion exchange 
in the $t$-channel was reggeized. It was found that for $W > 1.3$~GeV, the $Im[B_i(s,t)]$ in $\pi N$ electroproduction were well described by the 
resonant contributions only. This makes it possible to evaluate all Ball amplitudes from the resonance parameters by employing fixed-$t$ dispersion 
relations independently at different $Q^2$. For $W < 1.3$~GeV the non-resonant contributions, which become sizable in the Ball amplitudes, were 
evaluated by using the Watson theorem from the measured phases of the $\pi N$ elastic scattering amplitudes.

Within the UIM the full $\pi N$ electroproduction amplitudes are described by the resonant contributions and the non-resonant Born terms shown in
Fig.~\ref{npi_dynamics}~\cite{Aznauryan:2011qj,Aznauryan:2002gd}. The $s$- and $u$-channels incorporate nucleon exchange. The $t$-channel includes 
$\pi$, $\rho$, and $\omega$ exchanges. The reggeized $\pi$, $\rho$, $\omega$, $b_1$, and $a_2$ contributions are also employed in the description 
of the $t$-channel amplitudes. The full non-resonant amplitudes represent the sum of the Born pole terms and the reggeized $t$-channel contributions
with the relative weights fit to the data and detailed in Ref.~\cite{Aznauryan:2011qj}. The non-resonant contributions are unitarized within the 
$K$-matrix approach~\cite{Aznauryan:2002gd} assuming $\pi N$ rescattering for the final state interaction. The resonant contributions are described
within the Breit-Wigner ansatz with $W$-dependent $N^*$ electromagnetic, $\pi N$, and total decay widths. The electrocouplings were fit to the 
measured observables within the UIM approach. The uncertainties of the electrocouplings account both for the statistical/systematic data 
uncertainties, as well as the accuracy of the knowledge on the resonance hadronic decays and the electromagnetic and hadronic vertices in the 
non-resonant amplitudes.

\begin{figure*}
\centering
\includegraphics[width=0.90\textwidth]{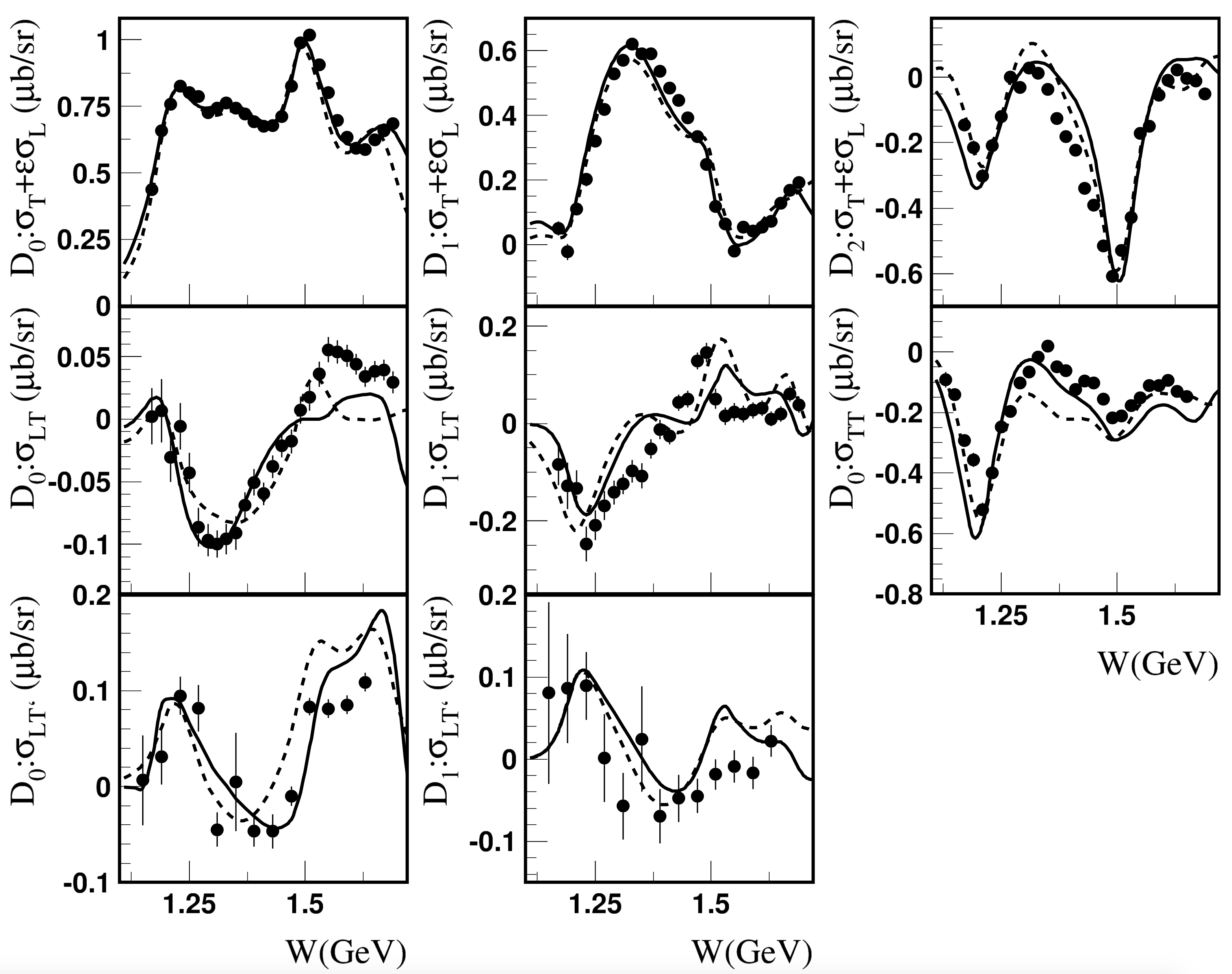}
\caption{Legendre moments of the $\pi^+n$ structure functions at $Q^2$=2.05~GeV$^2$. The experimental results from Ref.~\cite{Park:2007tn} are 
compared with the computations of the DR (solid lines) and UIM (dashed lines) approaches in Ref.~\cite{Aznauryan:2002gd}.}
\label{npi_legendre}
\end{figure*}

Both the DR and UIM approaches provide a good description of all $\pi N$ electroproduction data measured with CLAS at $W < 1.8$~GeV and
$0.16 < Q^2 < 5.0$~GeV$^2$. Representative examples are shown in Fig.~\ref{npi_integ} and Fig.~\ref{npi_legendre} for the description of the
fully integrated $\pi N$ cross sections and for the Legendre moments of the exclusive $\pi N$ structure functions, respectively. Differences in 
the electrocouplings determined within the DR and UIM approaches provide an estimate for the systematic uncertainties of these quantities 
determined from the different reaction models.  

\subsubsection{Electrocoupling Extraction from CLAS $\pi^+\pi^-p$ Electroproduction Data}
\label{phenom_2pi}

The analysis of the exclusive $\pi^+\pi^-p$ photo- and electroproduction data from CLAS was carried out within the data-driven meson-baryon 
JLab-Moscow (JM) model~\cite{Mokeev:2008iw,Mokeev:2012vsa}. The wealth of these data on the nine 1-fold differential $\pi^+\pi^-p$ cross sections 
allows us to establish all relevant mechanisms that contribute to this exclusive channel from their manifestations in the observables, such as 
peaks in the invariant mass distributions and pronounced dependencies in the CM angular distributions (see Fig.~\ref{exclusive_reactions2} (top)). 
The mechanisms with less pronounced kinematic dependencies can be pinned down from the correlations between their contributions in the different 
differential cross sections. Within this strategy, the phenomenological JM model has been developed with a primary objective to determine the 
resonance $\gamma_{r,v}pN^*$ photo-/electrocouplings and the resonance decay widths to the $\pi \Delta$ and $\rho p$ final states.

Within the JM model, the full 3-body $\pi^+\pi^-p$ electroproduction amplitude incorporates the contributions from five meson-baryon channels: 
$\pi^-\Delta^{++}$, $\rho p$, $\pi^+\Delta^0$, $\pi^+N^0(1520)3/2^-$, and $\pi^+N^0(1685)5/2^+$, with subsequent decays of the unstable intermediate 
hadrons. It also contains direct $2\pi$ photo-/electroproduction processes where the final $\pi^+\pi^-p$ state is created without the generation of 
unstable intermediate hadrons. Here the nucleon resonances contribute to the $\pi^-\Delta^{++}$, $\pi^+\Delta^0$, and $\rho p$ meson-baryon channels. 
Modeling of the non-resonant contributions is described in Refs.~\cite{Mokeev:2008iw,Mokeev:2012vsa,Burkert:2007kn,Ripani:2000va}. The model includes 
the contributions from all four-star PDG resonances with observed decays to the $\pi\pi N$ final states, the $N(1700)3/2^-$, and the new $N'(1720)3/2^+$
\cite{Mok20}. The resonant amplitudes are described within the unitarized Breit-Wigner ansatz~\cite{Mokeev:2012vsa} to make them consistent with the 
restrictions imposed by a general unitarity condition. The JM model offers a good description of the $\pi^+\pi^-p$ differential cross sections in 
the entire kinematic area covered by the data are shown in Table~\ref{pipip_kin_areas}. A representative example of the data description is shown in
Fig.~\ref{exclusive_reactions2}, along with the differential cross sections for the different contributing mechanisms inferred from the data. 

The resonance parameters were determined from a fit to the nine 1-fold differential cross sections under simultaneous variation of the
photo-/electrocouplings, masses, total and partial hadronic decay widths to the $\pi\Delta$ and $\rho p$ final states, and the non-resonant parameters 
of the JM model. The uncertainties of the resonant contributions are comparable with the uncertainties of the measured cross sections, which provides 
evidence for the reliable evaluation of these contributions and their photo-/electrocouplings.

\subsection{CLAS Electrocouplings from Exclusive Meson Electroproduction Data}
\label{nstar_electrocouplings}

The electrocouplings of $N^*$ states with masses below 1.8~GeV have been determined from analyses of CLAS $\pi N$, $\eta N$, and $\pi \pi N$ 
data for $Q^2$ up to 5~GeV$^2$. See Table~\ref{elcoupl_summary} for a summary of the results. A parameterization of these electrocouplings for 
this kinematic range is given in Ref.~\cite{HillerBlin:2019hhz}. 

\begin{table*}[htb]
\begin{center}
\vspace{2mm}
\begin{tabular}{|c|c|c|} \hline
Exclusive                  & Nucleon                        & $Q^2$ Range for  \\
Electroproduction Channel  & Resonance                      & $\gamma_vpN^*$ Electrocouplings         \\ \hline 
 $\pi^0 p$, $\pi^+n$       & $\Delta(1232)3/2^+$            & 0.16-6.0  \\ \cline{2-3}
                           & $N(1440)1/2^+$, $N(1520)3/2^-$, & 0.30-4.5  \\ 
                           & $N(1535)1/2^-$                &            \\ \hline
 $\pi^+n$                  & $N(1675)5/2^-$, $N(1680)5/2^+$ & 1.6-4.5  \\
                           & $N(1700)1/2^+$                 &      \\  \hline  
 $\eta p$                  & $N(1535)1/2^-$                 & 0.20-2.9  \\  \hline
 $\pi^+ \pi^- p$           & $N(1440)1/2^+$, $N(1520)3/2^-$, & 0.25-5.0  \\ \cline{2-3}
                           & $N(1520)3/2^-$, $\Delta(1620)1/2^-$, & 0.50-1.5  \\ 
                           & $N(1650)1/2^-$, $N(1680)5/2^+$, &            \\ 
                           & $\Delta(1700)3/2^-$, $N(1720)3/2^+$, $N'(1720)3/2^+$ &   \\  \hline
\end{tabular}
\end{center}
\caption{$N^*$ states for which the $\gamma_vpN^*$ electrocouplings are available from the CLAS exclusive meson electroproduction data showing 
the final state channel and the $Q^2$ range of the data.}
\label{elcoupl_summary}
\end{table*}

\begin{figure*}
\centering
\includegraphics[width=0.75\textwidth]{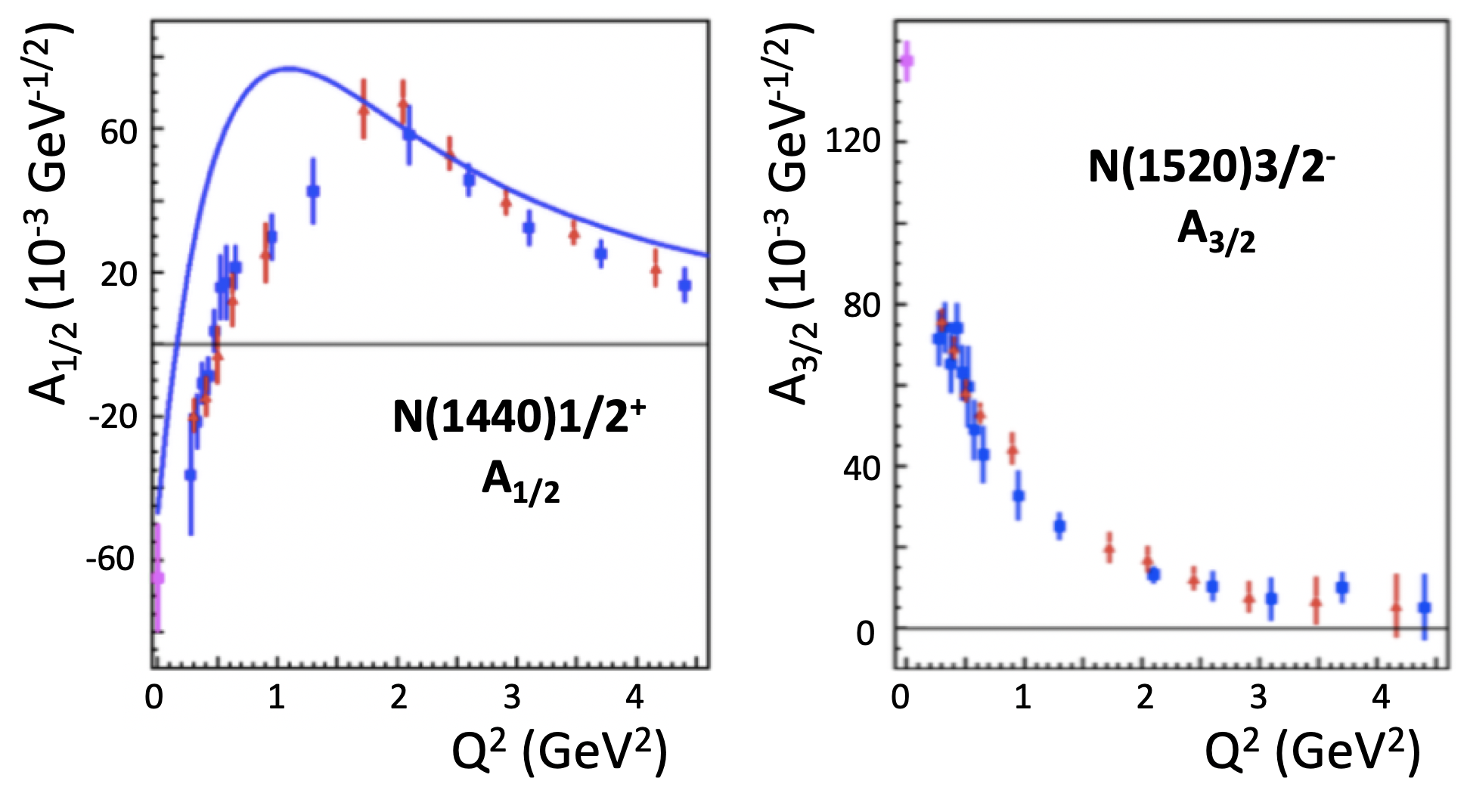}
\vspace{-1mm}
\caption{Electrocouplings of the $N(1440)1/2^+$ and $N(1520)3/2^-$ from independent analyses of the $\pi N$ (red triangles)~\cite{Aznauryan:2009mx} 
and $\pi^+\pi^-p$ electroproduction channels (blue squares)~\cite{Mokeev:2015lda,Mokeev:2019ron,Mokeev:2012vsa}. The $A_{1/2}(Q^2)$ electrocoupling 
for the $N(1440)1/2^+$ computed within the continuum QCD approach \cite{Segovia:2015hra,Rodriguez-Quintero:2019yec} is shown by the solid line. The 
region at $Q^2 > 1.5$~GeV$^2$ is the regime of quark core dominance.}
\label{npinpipi_elcoupl}
\end{figure*}

The most detailed information on the electrocouplings is currently available for the $N^*$ states in the mass range $W < 1.6$~GeV from 
studies of $\pi N$, $\eta N$, and $\pi^+\pi^-p$ photo-/electroproduction for $Q^2$ up to 7.7~GeV$^2$ for the $\Delta(1232)3/2^+$ and for $Q^2$ up 
to 7~GeV$^2$ for the $N(1535)1/2^-$~\cite{Aznauryan:2009mx,Denizli:2007tq,Villano:2009sn,Dalton:2008aa}, and for $Q^2 < 5$~GeV$^2$ for other
resonances~\cite{Aznauryan:2009mx}. The electrocouplings for the $N(1440)1/2^+$ and $N(1520)3/2^-$ available from independent studies of $\pi N$ 
and $\pi^+\pi^-p$ electroproduction are shown in Fig.~\ref{npinpipi_elcoupl}. These precise results for $\pi N$ and $\pi^+\pi^-p$ for $Q^2$ up to 
5~GeV$^2$ agree to within the data uncertainties for most data points. Consistent results on the electrocouplings from these two channels, which have 
entirely different non-resonant contributions, validates their extraction. Furthermore, this success demonstrates the capabilities of the reaction 
models discussed in Section~\ref{phenom} for the reliable extraction of the resonance parameters from independent studies of these two channels.

Studies of $\pi^+n$ and $\pi^+\pi^-p$ exclusive electroproduction have provided complementary information on the electrocouplings of the $N^*$ states 
in the third resonance region. The electrocouplings of the $N(1675)5/2^-$, $N(1680)5/2^+$, and $N(1700)1/2^+$ were determined for 
$1.6 < Q^2 < 4.5$~GeV$^2$ from $\pi^+n$ electroproduction~\cite{Park:2014yea}. For the first time, the electroexcitation amplitudes of the
$\Delta(1620)1/2^-$~\cite{Mokeev:2015lda}, $N(1720)3/2^+$, $N'(1720)3/2^+$, and $\Delta(1700)3/2^-$~\cite{Mok20}, which decay preferentially to the 
$\pi\pi N$ final state, have become available. Recently, new results on the exclusive structure functions for $\pi^0p$ electroproduction have also 
been obtained~\cite{Markov:2019fj}, together with the precise data from CLAS on the nine 1-fold differential $\pi^+\pi^-p$ cross sections
\cite{Fedotov:2018oan}. These observables have been obtained from the same data set measured for $1.4 < W < 1.85$~GeV and 
$0.4 < Q^2 < 1.0$~GeV$^2$ for both the $\pi N$ and $\pi^+\pi^-p$ channels. These data considerably extend the opportunities for the exploration 
of the electroexcitation amplitudes of the states in the third resonance region. 

A successful description of all of the CLAS nine 1-fold $\pi^+\pi^- p$ differential cross sections available in the range 
$1.4 < W < 2.0$~GeV and $2.0 < Q^2 < 5.0$~GeV$^2$~\cite{Isupov:2017lnd,Trivedi:2018rgo} has been achieved within the updated version of 
the JM model~\cite{Mokeev:2019ron,Burkert:2019opk}. This success offers good prospects for the extraction of the electrocouplings for most excited 
nucleon states over this entire kinematic range. 

These studies have made it apparent that consistency of the results from independent analyses of multiple exclusive channels is necessary to have 
confidence in the results. For the hadronic couplings, many high-lying $N^*$ states preferentially decay through the $\pi \pi N$ channel. For 
these states data from the $KY$ channels already measured with CLAS will be crucial to provide an independent analysis to compare the extracted 
electrocouplings for the high-lying $N^*$ states against those determined from the $\pi N$ and $\pi \pi N$ channels for $Q^2$ up to 5~GeV$^2$. 

\section{Insight into Strong QCD from Nucleon Resonance Electrocouplings}
\label{sec:2}

\subsection{Charting Active Degrees of Freedom in $N^*$ Structure}
\label{struct_components}

Detailed studies of the structure of most prominent nucleon resonances from the experimental results on their electroexcitation amplitudes have 
played a central role in the development of our understanding of how the strong interaction generates these states from quarks and gluons
\cite{Burkert:2019bhp,Aznauryan:2011qj,Roberts:2019wov,Aznauryan:2018okk}. The concept of the structure of $N^*$ states as bound systems of 
three constituent quarks that has emerged through such studies led to the development of the constituent quark model for nucleon resonances
\cite{isgur79,capstick86} (CQM) in the 1980s. As a result of intense experimental and theoretical effort over the past 30 years, it is now 
apparent that the structure of the nucleon and its spectrum of excited states is much more complex than what can be described in terms of the 
models based on constituent quarks alone~\cite{Burkert:2019bhp,Aznauryan:2011qj,Obukhovsky:2019xrs,Giannini:2015zia,Gutsche:2017lyu,Gutsche:2019yoo}. 

\begin{figure*}
\centering
\includegraphics[width=0.85\textwidth]{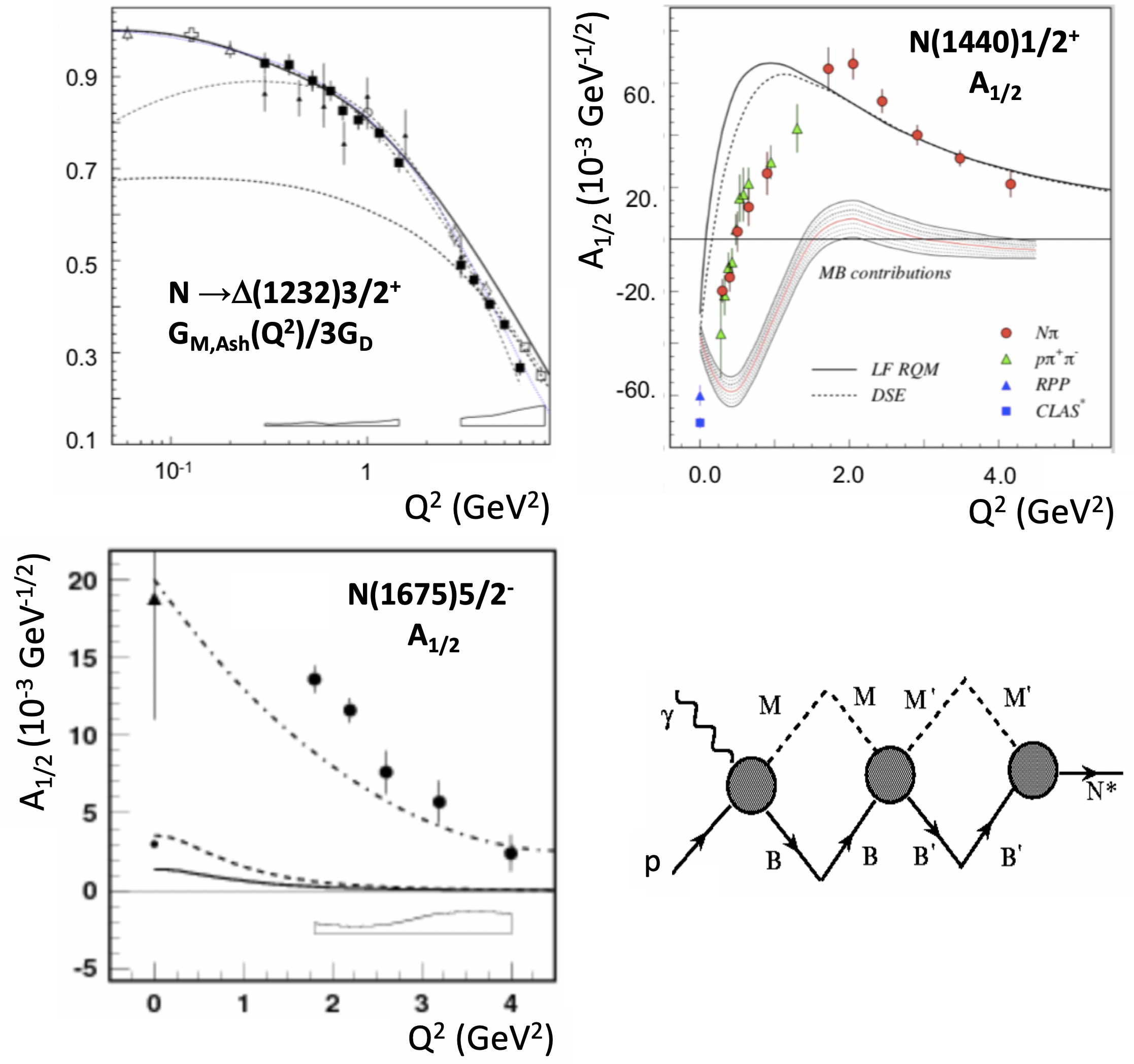}
\caption{Meson-baryon cloud and quark core in the structure of different resonances. (Top left) Data from the review of Ref.~\cite{Aznauryan:2011qj} 
on the $N \to \Delta$ magnetic transition form factor in the Ash convention $G^*_{M\,,Ash}$ normalized to the dipole fit of the ground state nucleon 
form factor $3G_D$ in comparison with model predictions~\cite{Suzuki:2010yn,Sato:2000jf} that account for the contributions from the combined quark 
core and meson-baryon cloud (solid line) and from the quark core only (thick dotted line). The evaluations from the large $N_C$ limit of QCD that 
incorporate a substantial part of the meson-baryon cloud~\cite{Pascalutsa:2006up} are shown by the dot-dashed and thin dotted lines. (Top right) 
CLAS results on the $A_{1/2}$ electrocouplings of the $N(1440)1/2^+$~\cite{Mokeev:2015lda,Aznauryan:2009mx,Mokeev:2012vsa} compared to the quark 
core contributions computed within the light-front relativistic quark model~\cite{Aznauryan:2012ec} (solid line) and within continuum QCD DSE
\cite{Segovia:2015hra} (dotted line). The meson-baryon cloud estimated as the difference between the experimental results and the light front 
quark model results for the quark core~\cite{Aznauryan:2012ec} is shown by the shadowed area. (Bottom left) $A_{1/2}$ electrocouplings of the 
$N(1675)5/2^-$ from CLAS data~\cite{Aznauryan:2014xea} compared to the contributions from the quark core evaluated within the hypercentral
\cite{Santopinto:2012nq} (solid line) and Bethe-Salpeter~\cite{Merten:2003iy} (dotted line) quark models. The meson-baryon cloud estimated within 
the Argonne-Osaka coupled-channel approach~\cite{JuliaDiaz:2007fa} is shown by the dot-dashed line. (Bottom right) The amplitudes that generate the 
meson-baryon dressing.} 
\label{delta_roper_n1675}
\end{figure*}

\begin{figure*}
\centering
\includegraphics[width=0.9\textwidth]{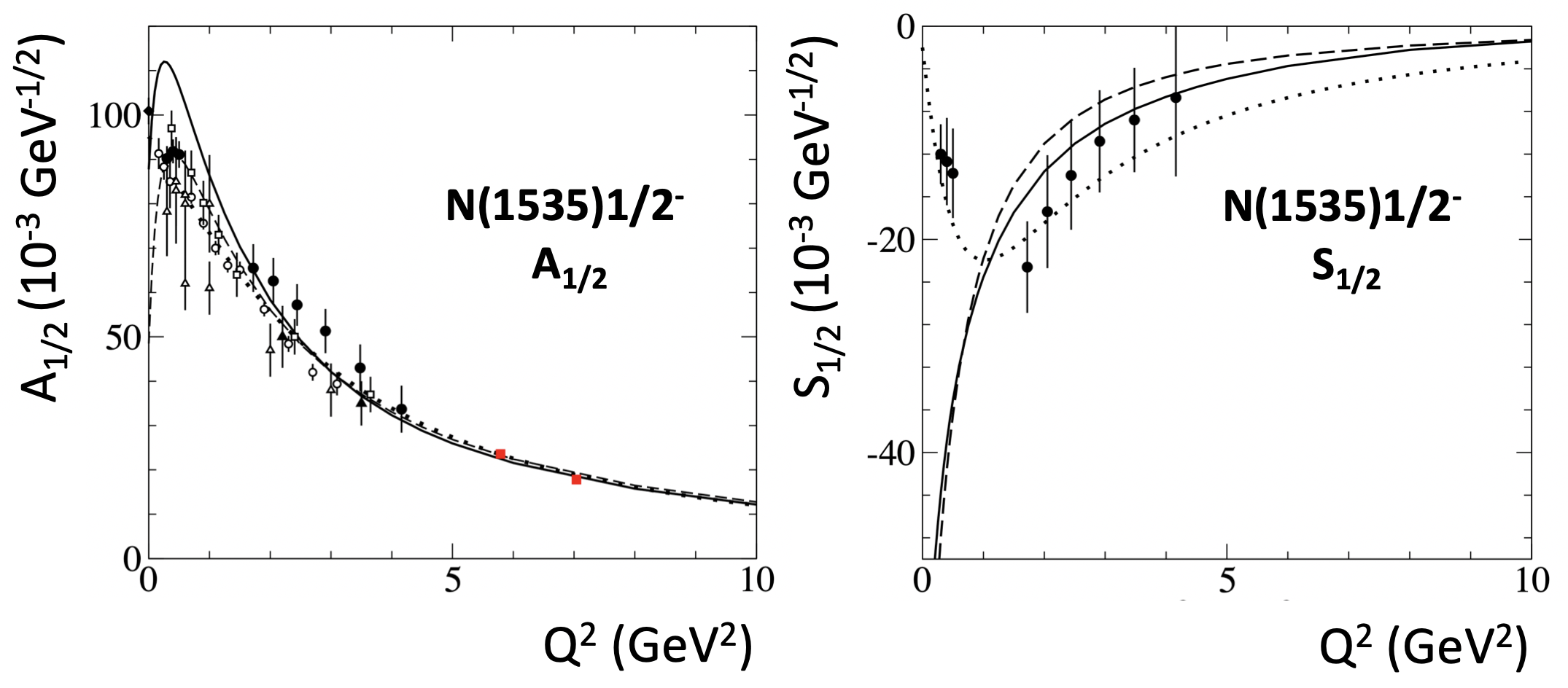}
\vspace{-2mm}
\caption{Improvements in the description of the results on the $\gamma_vpN^*$ electrocouplings~\cite{Aznauryan:2009mx,Dalton:2008aa} achieved after 
accounting for the contributions of the meson-baryon cloud and the quark core. A description of the $N(1535)1/2^-$ is shown for the relativistic quark
model~\cite{Obukhovsky:2019xrs} accounting for the contributions from $K\Lambda$ loops and the quark core (solid lines) and the quark core only (dashed 
line). The AdS/CFT results~\cite{Gutsche:2019yoo} are shown by the dotted lines.} 
\label{mb_cloud}
\end{figure*}

As shown in Fig.~\ref{delta_roper_n1675}, the approaches that take into account the contributions from dressed (constituent) quarks only are 
able to describe the experimental results on the electrocouplings for $Q^2 > 2$~GeV$^2$ (except for the $N(1675)5/2^-$), but they fail in reproducing 
the results for $Q^2 < 1$~GeV$^2$. The additional contributions to the structure of the states that are the most relevant at low $Q^2$ come from the 
meson-baryon cloud generated in the processes depicted in Fig.~\ref{delta_roper_n1675} (bottom right). For the $\Delta(1232)3/2^+$, the meson-baryon 
cloud contributions were inferred from the coupled-channel analyses of exclusive meson photo-, electro-, and hadroproduction data
\cite{Suzuki:2010yn,Sato:2000jf}. Only after accounting for the contributions from both the core of three dressed quarks and the meson-baryon cloud 
were the results reproduced for the $N \to \Delta(1232)3/2^+$ transition magnetic form factor in the entire $Q^2$-range covered by the measurements 
(see Fig.~\ref{delta_roper_n1675} (top left)). The continuum QCD Dyson Schwinger (DSE) approach offers a good description of the quark core contribution
to the $N(1440)1/2^+$ structure for $Q^2 > 2$~GeV$^2$ starting from the QCD Lagrangian~\cite{Segovia:2016zyc,Chen:2018nsg,Segovia:2015hra}. This success 
allowed for the evaluation of the meson-baryon cloud contributions as the difference between the experimental results on the $N(1440)1/2^+$ 
electroexcitation amplitude and the expectations from DSE on the quark core contributions. As shown in Fig.~\ref{delta_roper_n1675} (top right), 
the quark core contributions to the $N(1440)1/2^+$ $A_{1/2}$ electrocoupling evaluated within the light front quark model~\cite{Aznauryan:2012ec}, 
which employs the phenomenological quark mass function, are in good agreement with the DSE results~\cite{Segovia:2015hra} at $Q^2 > 2.0$~GeV$^2$. The 
meson-baryon cloud contributions to the $N(1440)1/2^+$ $A_{1/2}$ electrocoupling estimated as the difference between the experimental data and quark core 
contribution from the light front quark model \cite{Aznauryan:2012ec} is shown by shadowed area in Fig.~\ref{delta_roper_n1675} (top right). The 
meson-baryon cloud becomes the biggest contribution for $Q^2 < 1$~GeV$^2$, but almost vanishes for $Q^2 > 2$~GeV$^2$. The important role of the 
meson-baryon cloud in the structure of $N(1440)1/2^+$ was also demonstrated in recent lattice QCD studies that pave the way for the exploration of 
the emergence of the meson-baryon cloud~\cite{Sun:2019aem}. These studies have demonstrated that the structure of the $N(1440)1/2^+$ is defined by 
a core of three dressed quarks in their first radial excitation, surrounded by an external meson-baryon cloud, and they have resolved the more than 
fifty year old puzzle on the structure of this state~\cite{Burkert:2019bhp}. 

For other resonances studied with CLAS, the contributions from the meson-baryon cloud were taken into account within the model approaches
\cite{Obukhovsky:2019xrs,Aznauryan:2018okk,Gutsche:2017lyu,Gutsche:2019yoo,Aznauryan:2012ec}. Accounting for both dressed quarks and the meson-baryon 
cloud contributions offers an improved description of the electrocoupling results, in particular for $Q^2 < 1$~GeV$^2$ (see Fig.~\ref{mb_cloud}). The 
anti-de Sitter/conformal field theory (AdS/CFT) approach~\cite{Gutsche:2017lyu,Gutsche:2019yoo}, which effectively accounts for higher quark Fock space
configurations, provides a good description of the $N(1440)1/2^+$ and $N(1535)1/2^-$ electrocouplings over the entire $Q^2$ range, but with the most 
substantial improvements for low $Q^2$. This success emphasized the importance in $N^*$ structure of accounting for the meson baryon cloud generated 
by the higher quark Fock space configurations. The $N(1675)5/2^-$ resonance (see Fig.~\ref{delta_roper_n1675} (bottom left)) in the case of exact SU(6) 
symmetry and the single-quark transition electromagnetic current should not be excited in the $\gamma_v p$ reaction. SU(6) symmetry breaking 
results in a small value for the contributions from quarks as evaluated in Refs.~\cite{Santopinto:2012nq,Merten:2003iy}. These contributions are an 
order of magnitude smaller than the CLAS electrocoupling results. However, the implementation of the contributions from the meson-baryon cloud
\cite{Suzuki:2010yn,Sato:2000jf} allows for the description of the $N(1675)5/2^-$ electrocouplings. Therefore, these results for the first time probe 
the meson-baryon cloud almost completely. 

The structure of all $N^*$ states, as revealed by the studies of the electrocouplings, is determined by a complex interplay between the 
inner core of three dressed quarks and the external meson baryon cloud. The size of the meson-baryon dressing amplitudes is maximal for 
$Q^2 < 1$~GeV$^2$. As $Q^2$ increases there is a transition to the domain where the quark degrees of freedom just begin to dominate, as seen by 
the improved description of the $N^*$ electrocouplings obtained accounting for the quark core only (see Fig.~\ref{delta_roper_n1675} and 
Fig.~\ref{mb_cloud}) and also Refs.~\cite{Ramalho:2013mxa,Ramalho:2014pra}. For $Q^2 > 5$~GeV$^2$, the quark degrees of freedom are expected to fully 
dominate the structure of $N^*$ states~\cite{review}. Therefore, in the electrocoupling studies for $Q^2 > 5$~GeV$^2$ that are expected to become 
available for the first time in the $\pi^+\pi^-p$, $\pi N$, $\eta N$, and $KY$ channels with the data from CLAS12
\cite{E12-09-003,E12-06-108A,E12-16-010A}, the quark degrees of freedom will be probed more directly with only minor contributions from the 
meson-baryon cloud. This will mark the first opportunity to study experimentally this new and unexplored region of quark core dominance in the 
electroexcitation of nucleon resonances. For the foreseeable future, CLAS12 will be the only facility in the world capable of investigating $N^*$ 
structure at the distance scale where the dressed quarks emerge from QCD and generate the spectrum of excited nucleon states.

\subsection{Emergence of Hadron Mass from CLAS Data and New Opportunities with CLAS12}
\label{EHM}

Studies of the $\gamma_vpN^*$ electrocouplings shed light on the strong QCD dynamics that govern the generation of the dominant part of hadron
mass~\cite{Roberts:2018hpf,Segovia:2014aza,Segovia:2015hra,Mokeev:2015lda}. In resonance electroexcitation, the virtual photon interaction with 
the dressed quarks is sensitive to the dressed quark propagator. Hence, by varying $Q^2$, it become possible to map out the momentum dependence of 
the dressed quark mass through its manifestation in the $Q^2$-evolution of the electrocouplings.

\begin{figure*}
\centering
\includegraphics[width=0.9\textwidth,center]{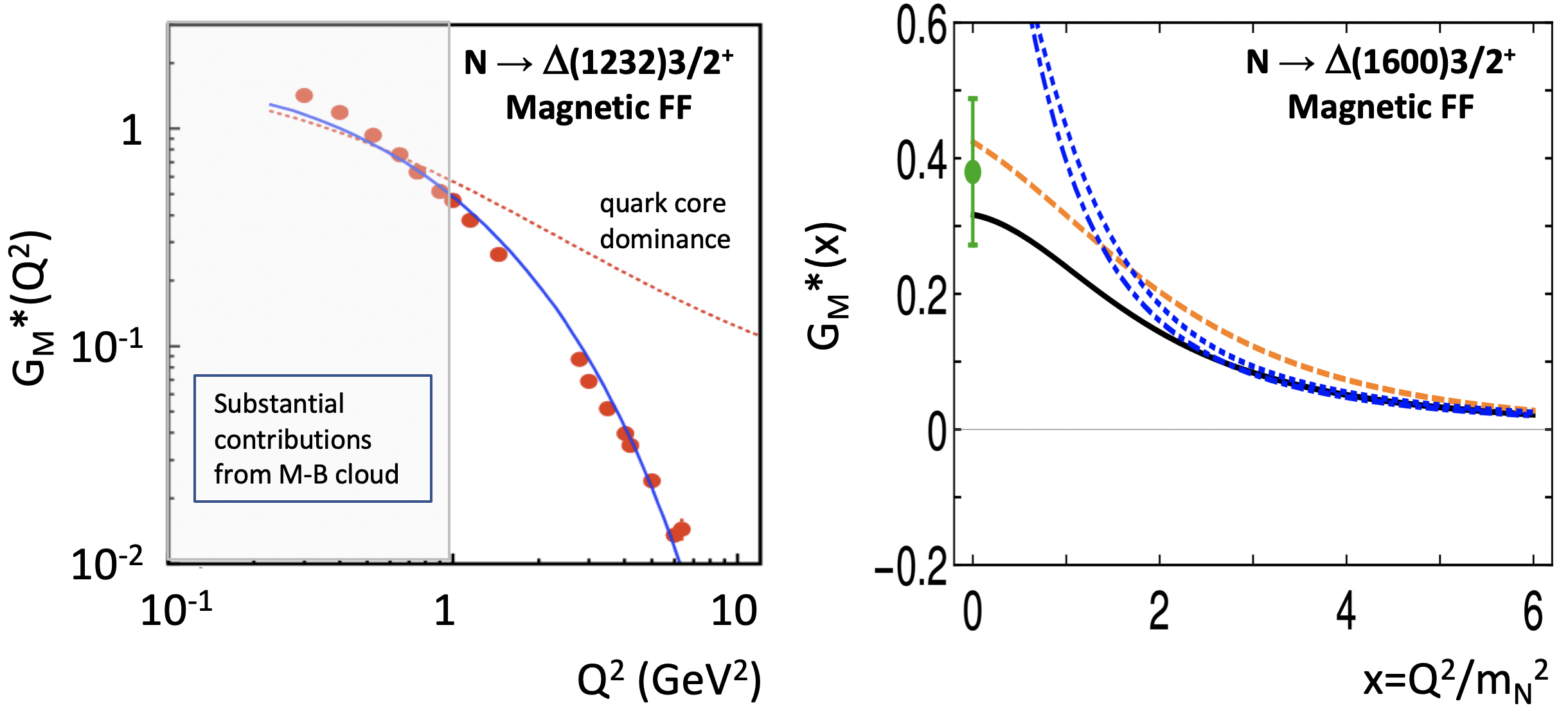}
\vspace{-4mm}
\caption{(Left) Description of the $N\to \Delta(1232)3/2^+$ magnetic transition form factor~\cite{Aznauryan:2009mx} with frozen (red dashed) and 
running (solid blue) quark mass~\cite{Segovia:2014aza}. (Right) Prediction for the $N\to \Delta(1600)3/2^+$ magnetic transition form factor within 
the continuum QCD approach~\cite{Segovia:2019jdk,Lu:2019bjs} with the same momentum dependence of the running mass of the dressed quarks as for the 
$\Delta(1232)3/2^+$, $N(1440)1/2^+$, and $\Delta(1600)3/2^+$ (black solid line). The DSE evaluations under the assumption that the wave function of 
the $\Delta(1600)3/2^+$ is reduced to the $s$-wave component only are shown by the blue lines. The dot-dashed line includes the full wave function 
of the proton and the dotted line assumes only the $s$-wave component. The DSE computation with an increased axial vector diquark contribution in 
the proton wave function is shown by the long-dashed orange line.} 
\label{ehm_deltap11}
\end{figure*}

The first evaluations of the $\Delta(1232)3/2^+$, $N(1440)1/2^+$, and $\Delta(1600)3/2^+$ electrocouplings starting from the QCD Lagrangian have 
become available from the continuum QCD DSE approach~\cite{Segovia:2014aza,Segovia:2016zyc,Chen:2018nsg,Segovia:2015hra,Segovia:2019jdk,Lu:2019bjs} 
for $Q^2 < 12$~GeV$^2$. As was discussed in Section~\ref{struct_components}, the resonance structure is defined by a complex interplay between the 
inner core of three dressed quarks and the external meson-baryon cloud. Currently, the DSE evaluations account for only the quark core component and 
can be directly confronted to the experimental results on the resonance electrocouplings only at high enough $Q^2$ where the contributions from the 
quark core play a major role.
 
The continuum QCD evaluations of the $N \to \Delta$ magnetic transition form factor and the $A_{1/2}$ electrocoupling of the $N(1440)1/2^+$ are 
shown in Fig.~\ref{npinpipi_elcoupl} (left) and Fig.~\ref{ehm_deltap11} (left). A good description of the experimental results has been achieved 
at $Q^2$ where the contributions from the quark core substantially exceed those from the meson-baryon cloud ($Q^2 > 0.8$~GeV$^2$ for the
$\Delta(1232)3/2^+$ and $Q^2 > 2$~GeV$^2$ for the $N(1440)1/2^+$). The continuum QCD evaluations of the $N \to \Delta$ magnetic transition form
factor~\cite{Segovia:2014aza} shown in Fig.~\ref{ehm_deltap11} (left) were carried out by using two different ans{\"a}tze for the $qq$-interaction.
In the initial exploratory computations, the results were obtained by using a simplified $qq$ contact interaction that accounts for the dynamical 
generation of the dressed quark mass of $\approx$350~MeV, but predicts the dressed quark mass independent of the quark momentum. The $N \to \Delta$ 
magnetic form factor computed with this frozen quark mass (shown in Fig.~\ref{ehm_deltap11} (left) by the dashed red line), overestimates the 
experimental results for $Q^2 > 1$~GeV$^2$, with the discrepancy increasing with $Q^2$. Eventually, this simplified description was replaced by the 
most advanced ansatz for the $qq$-interaction~\cite{kern1,kern2,kern3} that employs a dressed gluon propagator supported by the computations from the 
gauge sector of QCD~\cite{Binosi:2014aea}. The dressed quark propagator obtained as the solution of the gap equation by employing this advanced ansatz 
corresponds to a momentum-dependent (running) mass for the dressed quarks as shown in Fig.~\ref{mass_dse} (right). The continuum QCD evaluation, which 
incorporates the running mass of the dressed quarks, provides a good description of the experimental results on the $Q^2$-evolution of the $N \to \Delta$ 
magnetic transition form factor for $Q^2 > 0.8$~GeV$^2$ (shown in Fig.~\ref{ehm_deltap11} by the solid blue line). Analysis of the CLAS results on the 
$N \to \Delta$ magnetic transition form factor~\cite{Aznauryan:2009mx} within the continuum QCD approach~\cite{Segovia:2014aza} have conclusively 
demonstrated that the mass of the dressed quarks is running with the quark momentum.

Remarkably, a successful description of the $N(1440)1/2^+$ electroexcitation amplitudes for $Q^2 > 2$~GeV$^2$ (see Fig.~\ref{npinpipi_elcoupl} (left)) 
has been achieved with the dressed quark mass function evaluated under connection to the QCD Lagrangian that is {\em the same} as that used 
previously in the successful description of the pion~\cite{Horn:2016rip}, nucleon elastic, and $N \to \Delta$ transition magnetic form factors
\cite{Segovia:2014aza,Cui:2020rmu}. Therefore, the consistent results on the momentum dependence of the running dressed quark mass have become 
available from the independent studies of the pion and nucleon elastic form factors, as well as from analyses of the electroexcitation amplitudes of 
nucleon resonances of distinctively different structure, the $\Delta(1232)3/2^+$ (the spin-isospin flip of three dressed quarks) and the $N(1440)1/2^+$ 
(the first radial excitation of three dressed quarks). This success validates the relevance of dressed quarks with a momentum-dependent mass inferred 
from the QCD Lagrangian as the dynamical constituents in the structure of the ground state pion, the nucleon, and the lowest excited nucleon states. 
The capability to map out the momentum dependence of the dressed quark mass from analyses of the results on the $Q^2$-evolution of the nucleon elastic 
form factors and the electrocouplings has been conclusively demonstrated. A successful description of these quantities for prominent $N^*$ states with 
the same dressed quark mass function is critical in order to validate insight into this key ingredient of strong QCD. Insight into the strong 
QCD dynamics that control the generation of $>$98\% of the mass of hadrons from the results on the nucleon elastic form factor and the $\gamma_vpN^*$
electrocouplings represents one of the most impressive achievements in hadron physics of the last decade in the synergistic efforts between experiment,
phenomenology, and theory.

The continuum QCD approach~\cite{Segovia:2019jdk,Lu:2019bjs} has provided a parameter-free prediction on the $Q^2$-evolution of $N \to \Delta(1600)3/2^+$
$G^*_M(Q^2)$ magnetic transition form factor with the dressed quark mass function and all other elements in the Faddeev kernel for evaluation of the 
resonance mass and wave function exactly the same as used for the successful description of the electroexcitation amplitudes of the $\Delta(1232)3/2^+$ 
and $N(1440)1/2^+$ (see Fig.~\ref{ehm_deltap11} (right)). These predictions will be confronted with the experimental results from CLAS on the 
electroexcitation amplitudes of the $\Delta(1600)3/2^+$ that are expected in the near-term future. 

From the measurements with the CLAS12 detector outlined in Section~\ref{struct_components}, the results on the electrocouplings will be extended 
toward the highest $Q^2$ ever achieved in exclusive meson electroproduction data of $>5$~GeV$^2$. The expected quality of the results should be 
comparable to or better than those already available. The data expected from the experiments of the $N^*$ program with CLAS12 will make possible 
the study of the kinematic regime of quark momenta $0.5 < p < 1.1$~GeV (where $p = \sqrt{Q^2}/3$), running over the dressed-quark propagator (see
Fig.~\ref{mass_dse} (right)). The electrocouplings will be sensitive to the transition from the confinement regime of strongly bound dressed quarks 
and gluons at small momenta ($p < 0.5$~GeV) to the pQCD regime ($p > 2$~GeV), where almost undressed and weakly interacting current quarks and gauge 
gluons gradually emerge as the relevant degrees of freedom in the resonance structure with increasing $Q^2$. Consistent results on the dressed quark 
mass function from the independent analyses of the electrocouplings of all prominent nucleon resonances of different structure, including different 
spin-isospin flips, radial, and orbital three-quark excitations, will validate insight into this key ingredient of strong QCD.

Studies of the $N^*$ electroexcitations with CLAS12 will address key open problems in the Standard Model on the nature of hadron mass, its emergence 
from QCD, and its connection with DCSB. With the dressed quark mass function checked against the data on the nucleon elastic form factors and the
electrocouplings, the continuum QCD approach will be able to compute the parton pressure distributions in the nucleon ground state and to address 
another key open problem of the Standard Model on the emergence of quark-gluon confinement. A recent breakthrough in the studies of deeply virtual 
Compton scattering (DVCS) with CLAS~\cite{Burkert:2018bqq} has demonstrated the capability to gain insight into the parton pressure distribution. The 
accuracy of these results is expected to increase dramatically from the data of the DVCS experiments with CLAS12. 

\section{Conclusions and Outlook}
\label{concl}

Impressive progress has been achieved in the studies of nucleon resonance structure from the data on exclusive meson electroproduction. Such high-quality 
data from CLAS have allowed us to determine the electrocouplings of most resonances in the mass range up to 1.8~GeV with consistent results from analyses 
of the $\pi^+n$, $\pi^0p$, $\eta p$, and $\pi^+\pi^-p$ channels~\cite{Aznauryan:2011qj,Mokeev:2018zxt,Brodsky:2020vco}. This good agreement allows us to evaluate the 
systematic uncertainties for the electrocouplings related to the use of the reaction models. The analyses of the recent CLAS data on $\pi N$ and 
$\pi^+\pi^-p$ electroproduction~\cite{Markov:2019fj,Isupov:2017lnd,Trivedi:2018rgo} will further extend the information on the $\gamma_vpN^*$ 
electrocouplings for most $N^*$ states in the mass range up to 2~GeV for $0.2 < Q^2 < 5$~GeV$^2$ in the near-term future~\cite{Mokeev:2019ron,Burkert:2019opk}. 
The combined analyses of photo- and electroproduction data offer a promising new avenue in the exploration of the $N^*$ spectrum as was demonstrated in the 
recent observation of the new $N'(1720)3/2^+$ state from the $\pi^+\pi^-p$ data~\cite{Mok20}. The extension of the coupled-channel approaches successfully 
used for the search for missing resonances from the photoproduction data~\cite{Bur17,Kamano:2019gtm} toward the combined analyses of exclusive photo- and
electroproduction data is of particular importance in order to obtain the ultimate information on the spectrum of $N^*$ states generated in the strong QCD 
regime within the current decade. These developments will also facilitate the search for a new type of hybrid $N^*$ state with glue as an active structural
component in experiments with the CLAS12 detector~\cite{Burkert:2019kxy,E12-16-010}.

Analyses of the results on the $N^*$ electrocouplings~\cite{Aznauryan:2011qj,Mokeev:2018zxt,Mokeev:2015lda,Aznauryan:2009mx} within the continuum QCD
approach~\cite{Segovia:2014aza,Segovia:2016zyc,Chen:2018nsg,Segovia:2015hra}, quark models
\cite{Obukhovsky:2019xrs,Aznauryan:2018okk,Gutsche:2017lyu,Gutsche:2019yoo}, and the Argonne-Osaka global multi-channel framework
\cite{Kamano:2018sfb,Kamano:2019gtm} have revealed that the structure of all $N^*$ states studied with CLAS can be understood in terms of a complex 
interplay between an inner core of three dressed quarks and an external meson-baryon cloud. Future studies of the CLAS12 results on the nucleon 
resonance electrocouplings in the synergistic efforts between the continuum and lattice QCD approaches, and global coupled-channel analyses extended 
for the description of exclusive electroproduction, will shed light on the transition from confined dressed quarks in the resonance quark core to 
the meson-baryon cloud. These joint efforts between experiment and theory will address an important open 
problem on how quark-gluon confinement transforms into hadronic interactions between mesons and baryons.

The experimental results on the electrocouplings have provided important insights into the strong QCD dynamics that underlie the dominant part of 
hadron mass generation. The successful description of the CLAS results on the electrocouplings of the $\Delta(1232)3/2^+$ and $N(1440)1/2^+$
\cite{Mokeev:2015lda,Mokeev:2019ron,Aznauryan:2009mx,Mokeev:2012vsa} within the continuum QCD approach~\cite{Segovia:2014aza,Segovia:2015hra}, achieved 
with the {\it same} dressed quark mass function computed from the QCD Lagrangian, have demonstrated the capability to gain insight into this key ingredient 
of strong QCD. Furthermore, consistent results on the momentum dependence of the dressed quark mass from independent studies of the electrocouplings of 
nucleon resonances with distinctively different structure is critical in order to validate insight into the dynamics of hadron mass generation in a nearly
model-independent way. The continuum QCD approach~\cite{Segovia:2019jdk,Lu:2019bjs} has provided a parameter-free prediction for the electrocouplings of 
the $\Delta(1600)3/2^+$ by employing a universal quark mass function for the ground and excited nucleon states. The extraction of the resonance 
electrocouplings from the CLAS $\pi^+\pi^-p$ electroproduction data~\cite{Isupov:2017lnd,Trivedi:2018rgo} is in progress. Confirmation of the continuum 
QCD expectations on the electroexcitation amplitudes of this state will validate credible insight into the dynamics of hadron mass generation for
$Q^2 < 5$~GeV$^2$ covered by these data. Extension of the results on the electrocouplings toward $Q^2 > 5$~GeV$^2$ expected from the $N^*$ program with 
CLAS12 will allow us to map out the momentum dependence of the dressed quark mass at distances where the transition from quark-gluon confinement to pQCD 
is expected. These results will address a key open problem of the Standard Model on the emergence of hadron mass from QCD.

Theoretically, continuum QCD approaches, which employ a well-constrained dressed quark mass function and diquark correlation amplitudes mapped 
out in studies of the experimental results on the nucleon elastic form factors and the electrocouplings, are capable of evaluating the light-front wave 
function of the nucleon. This quantity can yield a complete theoretical description of the nucleon's shape in its intrinsic frame. Moreover, addressing
the emergence of the dynamical deformation seen in the structure of atomic nuclei, the continuum QCD approach can also describe the pion-exchange part of 
the $NN$-interaction with a $\pi NN$ vertex inferred from strong QCD within the same framework used for the description of the nucleon shape. Synergistic 
efforts in the experimental studies of the ground state nucleon and $N^*$ structure, combined with the developments in continuum QCD approaches and the 
theory of atomic nuclear structure~\cite{Launey:2016fvy,Dytrych:2018vkl}, pave the way toward understanding how the structure of atomic nuclei emerges 
from strong QCD \cite{Brodsky:2020vco}.

The dressed quark mass function that unifies the experimental results on the nucleon elastic form factors and the electrocouplings can also be used 
for the computation of the pion electromagnetic form factors and parton distribution functions (PDFs) within continuum QCD approaches. Hence, any 
conclusions drawn about the dynamics of hadron mass generation through studies of ground/excited-state nucleon structure can be validated via 
comparisons between the experimental results on the pion form factors/PDFs and the predictions from continuum QCD obtained with the same dressed quark
mass function~\cite{Brodsky:2020vco,Aguilar:2019teb,Horn:2016rip}.

The experiments in the 12~GeV era at Jefferson Lab with the CLAS12 detector, in conjunction with the completed and planned measurements with 
the datasets from the CLAS, offer a unique opportunity to map out the momentum dependence of the quark running mass at distances that correspond 
to the transition from bare QCD quarks to fully dressed constituent quarks. Insight into the dressed quark running mass from the experimental 
results on meson and baryon structure within the common framework offered by the continuum QCD approach will elucidate the emergence of $>$98\% 
of the hadron mass in the universe and how the structure of atomic nuclei is generated in the strong QCD regime. These objectives can be achieved 
in synergistic efforts between experiment, phenomenology, and the theory of hadron/nuclear structure with a traceable connection to QCD.

\begin{acknowledgements}
The authors gratefully acknowledge the fruitful discussions on these topics with Volker Burkert, Craig Roberts, and Jorge Segovia, each of whom 
provided valuable feedback during the preparation of this manuscript. This work was supported by the United States Department of Energy under JSA/DOE 
Contract DE-AC05-06OR23177. 
\end{acknowledgements}


%
%


\end{document}